\documentclass[fleqn,usenatbib]{mnras}

\usepackage{newtxtext,newtxmath}

\usepackage[T1]{fontenc}

\DeclareRobustCommand{\VAN}[3]{#2}
\let\VANthebibliography\thebibliography
\def\thebibliography{\DeclareRobustCommand{\VAN}[3]{##3}\VANthebibliography}

\usepackage{svg}
\usepackage{graphicx}	
\usepackage{amsmath}	
\usepackage{subcaption}
\usepackage{courier}

\newcommand{\corr}{\textcolor{black}}

\newcommand{\Mdot}{\mbox{\,$\mathrm{M}_{\odot}$}}        
\newcommand{\Zdot}{\mbox{\,$\mathrm{Z}_{\odot}$}}        
\newcommand{\ZLMC}{\mbox{\,$Z_{\rm LMC}$}}        
\newcommand{\Rdot}{\mbox{\,$ \mathrm{R}_{\odot}$}}        
\newcommand{\Ldot}{\mbox{\,$\mathrm{L}_{\odot}$}  }        
\newcommand{\aov}{$\alpha_{\rm ov}$}		
\newcommand{\fov}{$f_{\rm ov}$}		
\newcommand{\amlt}{$\alpha_{\rm MLT}$}		
\newcommand{\asemi}{$\alpha_{\rm semi}$}		

\title[Very Massive Stars]{The hydrogen clock to infer the upper stellar mass}
\author[E. R. Higgins et al.]{
Erin R. Higgins$^{1}$\thanks{E-mail: erin.higgins@armagh.ac.uk}, 
Jorick S. Vink$^{1}$,
Gautham N. Sabhahit$^{1}$\&
Andreas A.C. Sander$^{1,2}$\\
$^{1}$Armagh Observatory and Planetarium, College Hill, Armagh BT61 9DG, N. Ireland\\
$^{2}$Zentrum für Astronomie der Universit{\"a}t Heidelberg, Astronomisches Rechen-Institut, M{\"o}nchhofstr. 12-14, 69120 Heidelberg, Germany}

\date{Accepted 28 August 2022. Received 20 April 2022}
\pubyear{2022}

\begin{document}

\maketitle
\begin{abstract}
\noindent

The most massive stars dominate the chemical enrichment, mechanical and radiative feedback, and energy budget of their host environments. Yet how massive stars initially form and how they evolve throughout their lives
is \corr{ambiguous}. The mass loss of the most massive stars remains a key unknown in stellar physics, with consequences for stellar feedback and populations. In this work, we compare grids of very massive star (VMS) models with masses ranging from 80-1000\Mdot, for a range of input physics. We include enhanced winds close to the Eddington limit as a comparison to standard O-star winds, with consequences for present-day observations of $\sim$50-100\Mdot\ stars. We probe the relevant surface H abundances \corr{($X_{\rm s}$) }to determine the key traits of VMS evolution compared to O stars. We find fundamental differences in the behaviour of our models with the enhanced-wind prescription, \corr{with a convergence on the stellar mass at 1.6\,\corr{Myr}, regardless of the initial mass}. It turns out that $X_{\rm s}$ is an important tool in deciphering the initial mass due to the chemica\corr{lly} homogeneous nature of VMS above a mass threshold. We use $X_{\rm s}$ to break the degeneracy of the initial masses of both components of a detached binary, and a sample of WNh stars in the Tarantula nebula. We find that for some objects, the initial masses are unrestricted and\corr{, as such, }even initial masses of \corr{the} order 1000\Mdot\ are not excluded. Coupled with the mass turnover at 1.6\,\corr{Myr}, $X_{\rm s}$ can be used as a ‘clock’ to determine the upper stellar mass.

\end{abstract}


\section{Introduction}
The upper limit to the initial mass of stars remains a fundamental unknown in stellar astrophysics, with huge consequences for population synthesis, star formation theory and feedback studies \citep{Sobral15, Stanway15, Bouwens16, Berg18, Leitherer20}. Until recently, it was assumed that stars only formed up to masses of 100-150\Mdot\, \citep[e.g.][]{Fig05}. \corr{Observations and spectral modelling of massive stars in 30 Dor have, however, }provided evidence of stars with luminosities characteristic of current masses up to 200-300\Mdot \citep{Crow10, best14, martins15,Kalari22}. Moreover, the VLT-FLAMES Tarantula survey (VFTS) \corr{observed} an overabundance of massive stars in the Large Magellanic Cloud (LMC) with masses above 30\,\Mdot\, updating the Salpeter initial mass function (IMF) to a more top-heavy IMF \citep{Schneider18}. While we now have observational evidence for such massive stars above 150\Mdot, it remains unclear how massive stars initially form, and whether the upper mass limit remains the same across galaxies of varying metallicity ($Z$). Very massive stars \citep[{VMS, M$_{*}$ $\geq$ 100\Mdot;}][]{vinkbook} dominate the radiative and mechanical energy budget of their host galaxies, especially in young clusters where they dominate \citep{crow16}. Therefore, predicting the frequency and upper mass limit of VMS is critical for mapping their contributions \citep{Spera15,Graefener15,Roy20}. 

The self-enrichment of globular clusters (GCs) remains another key motivation for understanding the evolutionary history of VMS, as there currently exists a discord between theory and observations with respect to the dominant sources of enrichment, via asymptotic giant branch stars (AGB) and massive stars - known as the mass-budget problem \citep{bastianlardo18}.
The overabundance of VMS in the LMC might indicate a more dominant source of massive star self-enrichment in GCs in comparison to the AGB contribution, while the recently predicted slow winds of VMS, presented in a study of the effective upper mass \citep{vink18}, may allow these nuclear by-products \corr{(isotopes between Na and O, Mg and Al)} to remain inside the potential well of GCs.

Star formation of VMS remains highly debated with regards to the limiting factor of forming such massive protostars which do not fragment into multiple lower mass objects. Similarly, the formation and existence of supermassive stars (SMS, $M_{\rm{*}}$ $\geq$ 1000\Mdot) remains unknown, although they could be the products of mergers \citep[e.g.][]{gieles}. 
Ultimately, the effective upper mass limit of stars will differ across galaxies of varying $Z$, where lower $Z$ environments may be able to retain such high masses due to lowered wind mass loss than at Galactic $Z$. 

Stars with initial masses ($M_{\rm{init}}$) above $\sim$ 100\Mdot\, have very high luminosities which cause the radiative luminosity to approach the Eddington luminosity, leading to proximity to the Eddington limit ($\Gamma_{\mathrm{Edd}}$), where the ratio of radiative acceleration and gravitational acceleration become unity. \cite{GH08} and \cite{vink11} found that VMS which are close to their Eddington limit can produce strong optically thick winds which are radiatively driven by their high luminosities. Due to high $L/M$ values, they could produce strong Wolf-Rayet (WR) winds even during core H-burning \citep{dekoter97}. These stars are very massive WNh stars. 

Due to the relationship between convective core mass ($M_{\rm{cc}}$) and total mass ($M_{\rm{T}}$), VMS have extremely large convective cores which mix large amounts of nucleo-synthesised elements near the stellar surface \citep{hirschi15}. As a result, the evolution of VMS is fundamentally different to that of standard O-type stars. Where internal mixing dominates the evolution of 8-30\Mdot\, stars, and mass loss becomes important for the lives of stars in the range 30-60\Mdot, the most massive stars evolve chemically-homogeneously, implying that additional chemical mixing has a negligible \corr{impact} while strong mass loss drives the evolution towards a wide range of final masses as a function of $Z$. VMS are almost fully convective and as a result are very efficient energy generators which burn through their H with much shorter lifetimes than standard O-stars.

Moreover, as the $M_{\rm{cc}}$ remains such a large fraction \corr{($\sim$ 90\%)} of the stellar mass throughout the MS evolution, alongside radiative winds deteriorating the envelope mass, the star's surface composition is a proxy for the core evolution of VMS. As a result, the surface hydrogen (H) abundance can be utilised as an evolutionary `clock' during MS H-burning, while surface helium (He) enrichment can denote the proximity to the TAMS. In this study, we compare the surface \corr{($X_\mathrm{s}$) and core ($X_\mathrm{c}$) }H abundances with varying mass and internal mixing prescriptions to test its value as an evolutionary tracer of mass-loss histories and initial mass functions.

\cite{graef12} and \cite{San15} showed that when VMS evolve close to their Eddington limit, they produce H, He and iron (Fe) opacity bumps in the outer envelope. One solution to account for these opacity bumps is to balance the radiative pressure by creating radially extended low density envelopes which cause a density inversion close to the surface. These inflated envelopes may explain the large, empirically determined ``pseudo-photospheric'' radii of WR and LBV stars \citep[see for example, Fig. 5 in][]{Crow07}. However, with few observations of VMS to compare to, it remains unclear how stars deal with this proximity to $\Gamma_{\rm{Edd}}$ in Nature, whether their stellar winds sufficiently reduce their proximity to the Eddington limit by accounting for large amounts of mass being lost into the ISM, or by inflating their envelope to large radii in order to surpass the Fe opacity bump. Additional mixing in the stellar envelope may also quench this inflation by increased convection in the stellar envelope, similar to the superadiabatic nature of the \texttt{MLT++} routine \corr{which can be implemented to aid convergence} in MESA \citep{Pax15, Sabhahit21}. 

The most massive stars currently known are located in the R136 cluster of the LMC with masses predicted to lie in the 200-300\Mdot\ range. Their existence raises questions regarding their evolutionary and mass-loss history, and initial masses. We aim to test each of these properties and infer one from the other using updated mass-loss rates and insights from chemical homogeneous evolution (CHE), which is due to the large convective cores of VMS rather than rotation \citep{VinkHarries}. In our study, we address this by modelling the evolution of VMS with an updated wind prescription for a wide range of initial masses and a variety of internal mixing efficiencies. We provide details of our model grid in Sect. \ref{method}, alongside a sample of WNh stars and a detached binary from the Tarantula nebula, outlined in Sect. \ref{observations}. Results from our model grid are provided in Sect. \ref{results}. We introduce our discussion of the initial mass boundaries and qualitative differences in wind dependencies in Sect.\,\ref{boundaries}. Finally, we provide our conclusions in\corr{ Sect.} \ref{conclusions}.

\section{Method}\label{method}
In this work, we study the evolution of VMS \corr{through their main sequence}. We provide two methods of estimating the upper $M_{\mathrm{init}}$: an analytical calculation as a function of the mass-loss rate (Sect. \ref{boundaries}), and a direct comparison to stellar evolution models (see Sect.\,\ref{modelsSect}) and stellar atmosphere models. We test our methods with a larger sample of WNh stars from \cite{best14}. We also investigate the qualitative differences in implementing factors of the commonly-used \cite{Vink01} mass-loss rates with the newly-developed mass-loss prescription (Eq. \ref{mdotV11}) from \cite{Sabh22}, adopting the high Eddington mass-loss relations from \cite{vink11}. Subsequently, we evolve grids of stellar models for a range of convective and rotational mixing efficiencies.

\subsection{Stellar models}\label{models}
Our calculations for VMS models have been computed with the one-dimensional code \texttt{MESA} \citep[version 8845,][]{Pax11,Pax13,Pax15, Pax18, Pax19}. We developed a grid of models for masses 80-1000\Mdot\ in order to probe a wide range of possible initial masses. Our models are calculated for the Main-Sequence (MS) phase of evolution and are terminated upon H exhaustion, when the core-H abundance falls below $X_\mathrm{c}$\,$=$ 0.00001. The initial composition is based on the Large Magellanic Cloud (LMC) in line with our observations from the Tarantula Nebula \citep{best14, Shenar21}, at $\sim$50\% \Zdot, with \ZLMC\,$=$ 0.0088 and $Y$ $=$ 0.266. Relative scaled-solar abundances have been adopted from \cite{Grev98}, along with OPAL opacity tables from \cite{RogersNayfonov02} and the \corr{default nuclear network including 8 isotopes denoted \texttt{basic.net} in MESA \cite{Pax11}}. Nuclear reaction rates are taken from JINA REACLIB \citep{Cyburt10}. 

The Ledoux criterion for convection has been applied, with the standard mixing length theory (MLT) by \cite{Henyey64}, \corr{where a convective cell travels a length $l_{\rm{MLT}}$ set by a fraction \amlt\, of the pressure scale height ($H_{\mathrm{p}}$). We implement a standard value of} \amlt \,$=$ 1.5. Exponential overshooting has been implemented for core H-burning with values \fov\, $=$ 0.03 and 0.05, equivalent to \corr{a fraction of the pressure scale height \citep{Pax11}} \aov\, $=$ 0.3 for our standard reference models and 0.5 for high overshooting models. Superadiabatic mixing via the \texttt{MESA} parameter \texttt{MLT++} has been omitted from our standard models to test the effects of inflation with increased mass loss. Semiconvection is implemented with a \corr{diffusive efficiency denoted by the free parameter \asemi $=$ 100, \citep{L85}}. However, the inclusion of this process will not affect our results as the process requires a chemical gradient which only develops close to the TAMS and throughout core He-burning, lying beyond the investigations of this work. Standard spatial and timestep resolutions have been applied \texttt{varcontrol target} $=$ 10$^{-4}$ and \texttt{mesh delta} $=$ 0.3 \corr{\citep[see][]{MIST}}. The effects of rotation have also been tested with angular momentum transport and chemical mixing coefficients adopted from \cite{Heger00} for all rotating models. We calculate rotating models with an initial rotation rate set to $\Omega$/$\Omega_{\rm{crit}}$\, $=$ 0.4, corresponding to 40\%\, of the critical rotation rate which scales with $M_{\mathrm{init}}$. However, due to the inefficiency of rotation at the highest mass range caused by the strong outflows causing severe angular momentum loss, we focus on the effects of stellar winds and adopt non-rotating models for our standard grid in our comparisons. 

\subsubsection{Stellar winds}\label{windssect}
\corr{Theoretical studies of stellar winds have aimed to predict the amount of mass lost from the surface of stars for decades \citep{Cast75}. In \cite{Vink01}, theoretical mass-loss rates were calculated as a function of luminosity, mass, terminal velocity, effective temperature and metallicity. The resulting prescription was calculated for hot (log$_{10}$ (T$_{\mathrm{eff}}$/K) $\geq$ 4.0), optically thin OB stars, though studies have extrapolated this beyond canonical O stars to VMS \citep{yusof13,Koh15}. Empirical and theoretical studies \citep{vink06, best14} have shown that the standard O star wind prescription under-predicts mass loss of VMS. Moreover, uncertainties in the O star mass-loss rates have been estimated to be within a factor of 2 and 3 \citep{Bouret03,bjorklund21}. Subsequently, \cite{vink11} calculated mass-loss dependencies for MC simulations up to 300\Mdot\ finding a `kink' in the mass-loss rates with an increased dependency in L/M, which is in alignment with the observed spectral transition from O to Of/WNh stars. The predicted rates below the `kink' were also in good agreement with rates by \cite{Vink01}. Since the absolute mass-loss rates were not calculated in \cite{vink11}, a recent study by \cite{Sabh22} anchored the \cite{Vink01} rates to the observed transition point to provide a realistic mass-loss prescription which is applicable across a wide mass range (20-1000\Mdot). While the recipe outlined in \cite{Sabh22} is presented with dependencies in mass, luminosity, terminal velocity and metallicity, in this study we also include the temperature dependencies from \cite{Vink01} with the base rates accounting for this. We present the updated wind recipe for VMS  in eq.\eqref{mdotV11} with comparable dependencies formatted as seen in eq. (24) of \cite{Vink01}.}

The default stellar wind prescription outlined in the \texttt{MESA} code is called the `Dutch' wind scheme, which incorporates the \cite{Vink01} rates during core H-burning, and the \cite{NugisLamers} rates when $\mathrm{X_{s}} < 0.4$, during the hot wind regime. The cool supergiant regime adopts the \cite{deJager} mass-loss rates, though \cite{Vink21} showcased how the chosen effective temperature \corr{(log$_{10} (T_{\mathrm{eff}}$/\rm{K}) $\sim$ 4.0-3.6)} in the switch from one recipe to another can affect the final mass and potential fate of massive stars on their way to becoming black holes or pair instability supernovae. In this work, we \corr{test multiplications of the base mass-loss rates of \cite{Vink01} for O stars, increased by a factor of 2 and 3. We then compare these factors ($f_{\mathrm{V01}}$ $=$ 1, 2, 3) of the \cite{Vink01} mass-loss rates} with that of the updated \cite{vink11} rates, outlined in \cite{Sabh22}. 

Theoretical studies such as \cite{vink06}, \cite{GH08} and \cite{vink11} have provided key insights into the behaviour of winds driven by VMS, mainly in their dependencies on $M_{\mathrm{init}}$, metallicity and proximity to the Eddington limit, where the total Eddington parameter is defined as\corr{,
\begin{equation}
    \Gamma = \frac{\kappa L_{\rm{rad}}}{4 \pi cGM}. 
\end{equation}
}In this total $\Gamma$, $\kappa$ denotes the total, flux-weighted mean opacity which increases dramatically in VMS envelopes due to the Fe-opacity peak, \corr{which results in} $\Gamma$ approaching unity. Alternatively, one can define the `classical' Eddington parameter $\Gamma_{\rm{e}}$ related to the Thompson opacity $\kappa_{\rm{e}}$ due to free electron scattering. For a fully ionised plasma, this parameter can be denoted as\corr{,
\begin{equation}
    \Gamma_{\rm{e}} = 10^{-4.813} (1+X_{\rm s})\frac{L}{M}. 
\end{equation}
}\cite{GH08} investigated the dependencies of VMS winds when approaching the Eddington limit, finding a strong dependence on $Z$ and $\Gamma_{\rm{e}}$. \cite{vink11} found a `kink' in the mass-loss rates of massive stars with Monte Carlo radiative transfer models of masses ranging 40-300\Mdot\ due to a mass-loss transition \citep{VG12} when $\Gamma_{\rm{e}}$ exceeds a critical value. While it is clear that the slope of VMS winds is steeper than that of O stars, the exact dependence on total $\Gamma$ or $\Gamma_{\rm{e}}$ has remained highly uncertain.

\cite{Sabh22} investigated the effects of implementing a pure $L/M$-dependence in the winds of stellar models compared to a steep $\Gamma_{\rm{e}}$ dependence which accounts for a change in $X_{\rm s}$. They found that when comparing to observations of VMS in the Arches cluster and 30 Dor cluster of the LMC, models which include a strong (1+$X_{\rm s}$) \corr{exponent} have difficulties in reproducing the effective temperature range of the VMS in both galaxies. However, models including a pure $L/M$-dependence, have a self-regulatory \corr{behaviour} leading to an effective temperature range which agrees with both sets of observations due to the \corr{significant} drop in luminosity leading to a narrow range in $T$. It is important to note that while the true $\Gamma_{\rm{e}}$ scaling may have some $X_{\rm s}$ dependence, tests by \cite{Sabh22} already show that a strong $X_{\rm s}$ dependence is very sensitive \corr{in balancing the effects of envelope inflation and mass loss} while a weak $X_{\rm s}$ dependence produces the self-regulatory \corr{evolution} which can reproduce the effective temperature range of VMS observations.

\subsubsection{Updated wind prescription for VMS}
Empirical mass-loss rates of Of/WNh transition stars have allowed for detailed studies of the `kink' in mass-loss rate at higher masses (60-100\Mdot). Observations in the LMC have provided accurate transition mass-loss rates, with corresponding transition masses and luminosities. With that, the \cite{Vink01} rates can be extended for the high $\Gamma$ exponents and anchored at the observed transition region or `kink', where the resulting mass-loss recipe is: \begin{equation}\label{mdotV11}
\begin{split}
    \mathrm{log\corr{_{10}}} \; \dot{M} = \;\;& -8.441 \\ &
    + 4.77\;\mathrm{log\corr{_{10}}}\;(L/\rm{L}_\odot/10^5) \\ &
    - 3.99\; \mathrm{log\corr{_{10}}}(M/\rm{M}_\odot/30) \\ &
    - 1.226\; \mathrm{log\corr{_{10}}} (\varv_{\rm{inf}}/\varv_{\rm{esc}}/2) \\ &
    + 0.933 \mathrm{log\corr{_{10}}} (T_{\mathrm{eff}}/K/40000)\\ & 
    - 10.92\; \mathrm{log\corr{_{10}}}{(T_{\mathrm{eff}}/K/40000)}^2\\ &
    + 0.85\;\mathrm{log\corr{_{10}}} (Z/\rm{Z}_\odot).\\
\end{split}
\end{equation}
In this work, we explore the $L/M$-dependent wind prescription adopted from \citet[][]{vink11} and \cite{Sabh22} which allows models to remain in stable evolution below the Eddington limit, at least through the MS evolution.

\begin{table}
    \centering
    \begin{tabular}{|c|c|c|c|c|}
    \hline
        Star & Spectral type &  $\mathrm{log\corr{_{10}}}\;(L/\rm{L}_\odot)$ & $\mathrm{log\corr{_{10}}} \, (T_{\mathrm{eff}}$/K) & $Y_{\mathrm{He}}$ \\
        \hline \hline
        VFTS1025 & WN5h & 6.58 & 4.63 & 0.7 \\
        VFTS758 & WN5h & 6.36 & 4.67 & 0.775\\
        VFTS695 & WN6h + ? & 6.50 & 4.60 & 0.85\\
        VFTS427 & WN8(h) & 6.13 & 4.60 & 0.925\\
        VFTS108 & WN7h & 5.7 & 4.60 & 0.775\\
        \hline
    \end{tabular}
    \caption{Stellar parameters of all WNh stars from the VLT-FLAMES Tarantula Survey sample with surface He abundances $\geq$ 70$\%$, adopted from \citet{best14}.}
    \label{tab:VFTS}
\end{table}

\begin{table}
    \centering
    \begin{tabular}{|c|c|c|c|c|}
    \hline
        $M_{\mathrm{dyn}}/\mathrm{M}_{\odot}$ &  $\mathrm{log\corr{_{10}}}\;(L/\rm{L}_\odot)$ & $\mathrm{log\corr{_{10}}} \, (T_{\mathrm{eff}}$/K) & $X_{\rm s}$ & $\mathrm{log\corr{_{10}}}\;(R/\rm{R}_\odot)$  \\
        \hline \hline
        74 $\pm$ 4 & 6.44$\pm$ 0.05 & 4.65 & 0.35$\pm$ 0.05 & 1.4\\
        69 $\pm$ 4 & 6.39 $\pm$ 0.05 & 4.60 & 0.4$\pm$ 0.05 & 1.5\\
        \hline
    \end{tabular}
    \caption{Stellar parameters of the detached eclipsing binary R144, adapted from spectral analysis by \citet{Shenar21}. Dynamical masses, luminosities, effective temperatures, $X_{\rm s}$ abundances and stellar radii are included.}
    \label{tab:R144}
\end{table}

\subsection{Observational sample of VMS in the LMC}\label{observations}
Due to the shape of the initial mass function (IMF), massive stars are increasingly rare \corr{for higher $M_{\mathrm{init}}$, resulting in fewer observations at the upper mass range}. However, due to surveys such as VLT-FLAMES \citep{Evans05} and the VLT-FLAMES Tarantula Survey \cite[VFTS][]{Evans11}, hundreds of massive star spectra are available giving an insight into the evolutionary traits and populations of O and B stars. \corr{However, with fewer observations of VMS compared to O and B stars}, it is increasingly challenging to study the lives of VMS. Consequently, our understanding of VMS relies on theoretical studies such as \cite{vink11} \corr{and} \cite{vink18}. The Arches cluster, however, is a young star-forming region near the Galactic centre which hosts VMS \citep{martins08}. This cluster contains 13 very luminous WNh stars which have been utilised in \cite{Graefener+2011} to determine mass-luminosity relations for VMS and wind dependencies as a result. These WNh stars were also utilised in the recent study by \cite{Sabh22} in testing the $\Gamma_{\rm{e}}$-dependence of VMS winds.

\subsubsection{WNh stars in VFTS}
In this study, we consider observed WNh stars as a comparison to our theoretical models of VMS. We select the WNh observations from the VFTS sample \citep{best14} which have high luminosities representative of VMS evolution, and low $X_{\rm s}$ abundances. Since H is observed in the spectra, we can assume that there is still a small remainder of \corr{H on the surface} while having a surface He abundance greater than 70\%. Table \ref{tab:VFTS} provides a list of all the WNh objects from the VFTS which have surface He abundances $\geq$ 70\%, leaving  $\sim$ 5-30\% $X_{\rm s}$. While the effective temperatures are uncertain, they all lie within 0.1 dex.

\subsubsection{R144: the most massive binary in TMBM}
Following the VFTS studies of massive stars, a multi-epoch campaign was launched to survey 100 massive binary candidates known as the Tarantula Massive Binary Monitoring \citep[TMBM,][]{TMBM1, TMBM2, TMBM3, TMBM4}. \cite{Shenar21} provided the analysis of the most massive binary in the TMBM sample, R144, in Paper V. With compact radii \corr{(log$_{\rm{10}}$($R/\Rdot$) $=$ 1.4)} and high Eddington parameters \corr{($\sim$0.78)}, coupled with high luminosities \corr{(log$_{\rm{10}}$($L/\Ldot$) $=$ 6.4)} and low $X_{\rm s}$ abundances \corr{($X_{\rm s}$ $=$ 0.35)}, it was unclear whether the components of R144 were core H-burning or He-burning objects. If the objects are core H-burning, one would expect their radii to be significantly more inflated than observed as the stars are very close to the Eddington limit. With compact radii and moderate dynamical masses for such high luminosities on the other hand, we would expect the objects to be core He-burning WR stars. Still, the combination of derived effective temperatures and luminosities places them towards the MS. Hence, the primary and secondary components of R144 provide an interesting study of the lives of VMS and what observational traits could be most reliable in breaking the degeneracies in their possible evolutionary histories.
\begin{figure}
    \includegraphics[width = \columnwidth]{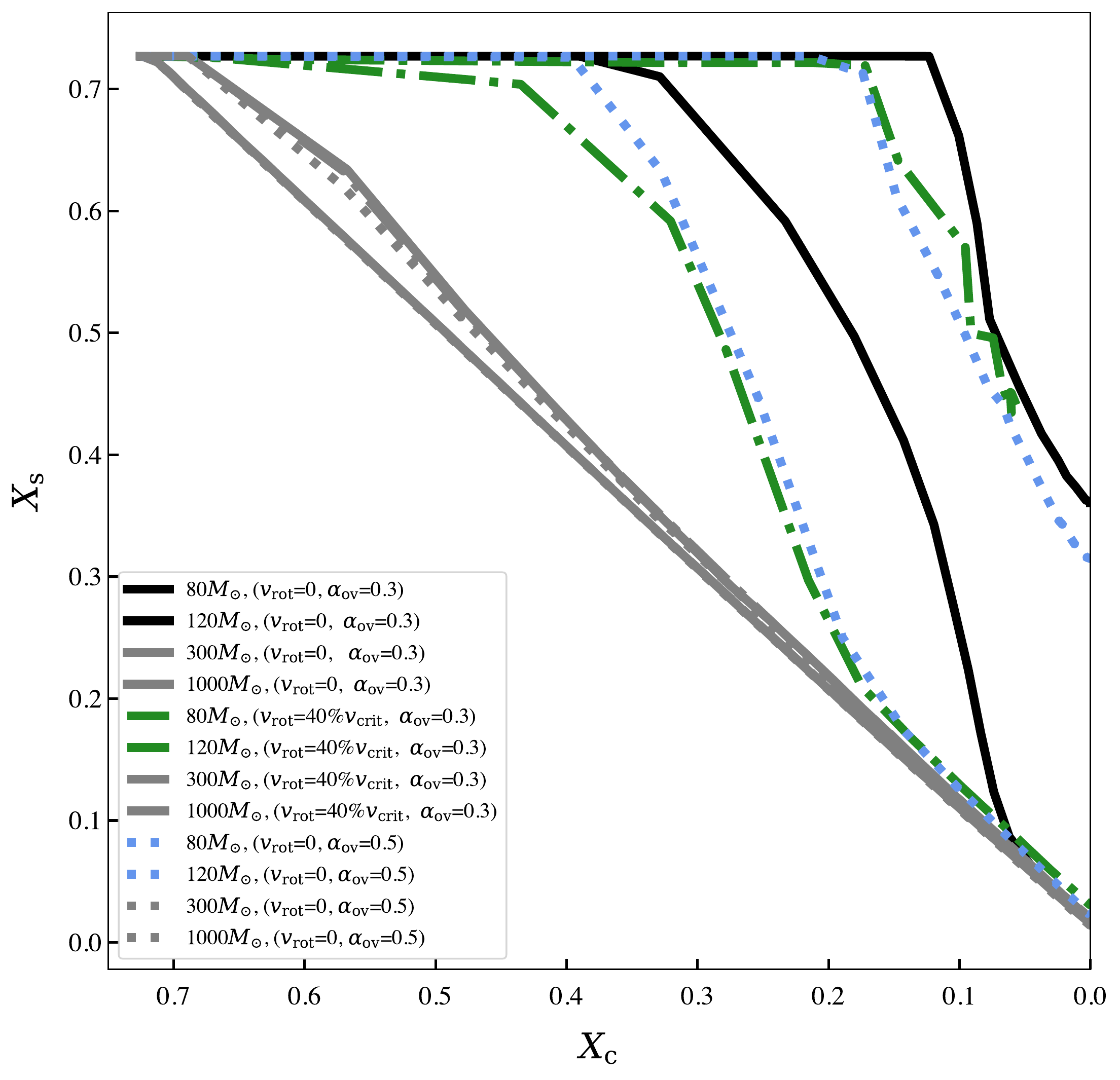}
    \caption{Surface H \corr{($X_{\rm s}$)} abundance as a function of core H ($X_{\rm c}$) abundance during MS evolution of models with $M_\mathrm{init}$ $=$ \corr{80\Mdot, 120\Mdot, 300\Mdot\ and 1000\Mdot. Dotted blue lines represent models with \aov\,$=$ 0.5 (rather than 0.3 in the standard models shown by solid black lines). Dash-dotted green lines show rotating models with 40\% critical rotation, rather than the standard non-rotating models shown by solid black lines. Grey tracks illustrate models with $M_\mathrm{init}$ $=$ 300\Mdot\ and 1000\Mdot\ }for our complete range of input physics including \aov\,$=$ 0.3 and 0.5, with 0\% and 40\% critical rotation.}
    \label{fig:Xsurf-Xcore}
\end{figure}

\section{Results}\label{results}
We have calculated model grids of stellar masses 80-1000\Mdot\ with updated wind dependencies from \cite{vink11}, and for three sets of internal mixing prescriptions. Our models demonstrate that in the VMS range above $\sim$ 200\Mdot, the effects of \corr{increased chemical mixing via convective core overshooting or rotation} have little effect on the trajectory of the evolution of these models. This is due to the chemical homogeneity seen in Fig.\,\ref{fig:Xsurf-Xcore}, where 80-120\Mdot\ models \corr{(green and blue dashed lines)} are distinguishable depending on their internal mixing, whereas 300-1000\Mdot\ models \corr{(grey tracks)} show uniform evolution irrespective of mixing prescription. \corr{In this study, we compare grids of rotating and non-rotating models, as well as models with varied core overshooting \aov. Since VMS have large convective cores ($\sim$90\% of the total stellar mass), the extension of the core by the process of convective overshooting, whereby convective cells travel beyond the convective boundary due to their non-zero velocity, has a negligible effect. The $H_{\mathrm{p}}$ is reduced at extended radii, and as such the fractional extension by \aov\ is less significant than for a 30\Mdot\ star. Similarly, the effect of rotation on internal mixing and angular momentum transport is reduced due to the increased effect of stellar winds at such high $M_\mathrm{init}$ (100-1000\Mdot). Hence, for the purpose of studying VMS evolution, we adopt non-rotating models with moderate overshooting (\aov\, $=$ 0.3) as our standard grid of models, while comparing the effects of additional mixing via rotation (40\% critical) and increased overshooting (\aov\, $=$ 0.5).}

\subsection{Surface \corr{h}ydrogen abundance as a proxy for core evolution}

Figure \ref{fig:Xsurf-Xcore} demonstrates that above $\sim$200\Mdot, stars experience so-called chemically-homogeneous evolution (CHE). Consequently, the surface abundances reflect the core fusion products and central abundances. The models shown in Fig.\,\ref{fig:Xsurf-Xcore} demonstrate that CHE is mass dependent and that rotation only plays a minor role, considering that below 200\Mdot\ (green dash-dotted lines) models do not produce fully mixed stars when including rotation but instead above $\sim$ 200\Mdot\ stars show a qualitatively different core-surface behaviour. Models in the 80-120\Mdot\ mass range can alter their surface enrichments based on the amount of internal mixing included via rotation or convective overshooting, also shown in Fig. \ref{fig:Xsurf-Xcore}. VMS \corr{however, undergo CHE} due to their large convective cores which enclose $\sim$ 90\% of their total mass \citep[][]{yusof13, hirschi15, Kohler15}, making additional mixing by rotation or overshooting ineffective. \corr{In fact, an increase in the $H_{\mathrm{p}}$ of VMS relates to a smaller relative core increase than at higher masses.} This means that \aov$=$ 0.5 does not have the same effect in a 500\Mdot\ model as in 50\Mdot\ models. 

If we consider the CHE of VMS, we can expect observations with high $X_{\rm s}$ abundances to be in the first half of their MS, while low $X_{\rm s}$ abundances suggest that these stars are towards the end of their MS lifetime. Figure \ref{fig:Xsurf-Xcore} shows that lower mass models $M_\mathrm{init}$ $\sim$ 80-120\Mdot, with increased mixing also have surface abundances similar to VMS (grey) tracks in the last $\sim$ 20\% of their MS lifetimes. This means that surface abundances of $\sim$ 20\% H correlate to $X_{\rm c}$ of 20\% for both VMS and lower mass models with additional mixing. Now, in the case of our sample of WNh stars from \cite{best14}, we observe a surface He mass fraction of 70-92.5\% in all cases, which could suggest that these stars either originate from stars in the 80-150\Mdot\ range and could be core He-burning, H-depleted objects. Alternatively, they could also originate from core H-burning VMS with $M_\mathrm{init}$ above $\sim$ 300\Mdot.

If the stars in Table \ref{tab:VFTS} were core He-burning, then high mass-loss rates would quickly remove the last 10-20\% \corr{of H at the surface} and would \corr{result in $X_{\rm s}$ reaching zero shortly after beginning core He-burning.} VMS (M$>$100\Mdot) with high mass-loss rates lead to reduced $X_{\rm s}$ during the MS, but \corr{an }$X_{\rm s}$ which falls below $\sim$ 30\% would suggest that the even steeper WR winds during core He-burning would further deplete the star of $X_{\rm s}$. Therefore, even though the final 10\% of H-burning and the entire He-burning lifetime equates to $\sim$ 100,000 years, the core He-burning winds would lead to significant \corr{loss of H-rich material} at the surface, \corr{compared with the final 10\% of core H-burning where the mass-loss rates are lower}. This means that if an object is observed with a high luminosity \corr{ ( $\mathrm{log\corr{_{10}}}\;(L/\rm{L}_\odot)$ $\geq$ 6.0)} and $X_{\rm s}$ $=$ 0-20\%, it is likely core H-burning. From Fig.\,\ref{fig:MH}, we further learn that the $M_\mathrm{init}$ of a luminous WNh star is essentially unknown if the $X_{\rm s}$\, abundance is below $\sim$30\%. \corr{These stars could either originate from a moderate 100-200\Mdot\ star with increased mixing or a VMS with $M_\mathrm{init}$ $=$ 300-1000\Mdot. At these observed high luminosities and masses, both objects are in the final 10-30\% of their MS, indistinguishable in their HRD position and mass\corr{-loss} history.}

\subsection{\corr{Importance of stellar winds on the upper stellar mass}}\label{IMFwinds}
As outlined in Sect.\,\ref{windssect}, current stellar evolution codes employ a range of mass-loss prescriptions for various phases of evolution, with the commonly-used \texttt{DUTCH} configuration used in \texttt{MESA} for massive stars evolving with hot star winds which utilise the rates from \cite{Vink01}. In stars with masses above the \corr{canonical} O-star regime \corr{($\gtrsim$ 60\Mdot)}, it is unclear which mass-loss rates are most appropriate, with some studies varying the \cite{Vink01} rates by a range of factors. \cite{vink11} highlighted a strong $\Gamma$ dependency in their theoretical models, with other suggestions by \cite{best20} \corr{and} \cite{graef21}. In this work, we compare the effects of the standard \cite{Vink01} winds with increased rates by a factor of \corr{ 2 and 3 }for models with $M_\mathrm{init}$ $=$ 80-1000\Mdot. Following this, we then compare with the $L/M$ wind prescription detailed in \cite{Sabh22} which does not include a strong (1+$X_{\rm s}$) \corr{exponent}, as discussed in Sect.\,\ref{windssect}. Instead\corr{,} this updated prescription adopts M and L \corr{exponent}s from \cite{vink11} to provide an appropriate enhanced wind mass loss, see Eq.\,\eqref{mdotV11}.

\begin{figure}
    \includegraphics[width = \columnwidth]{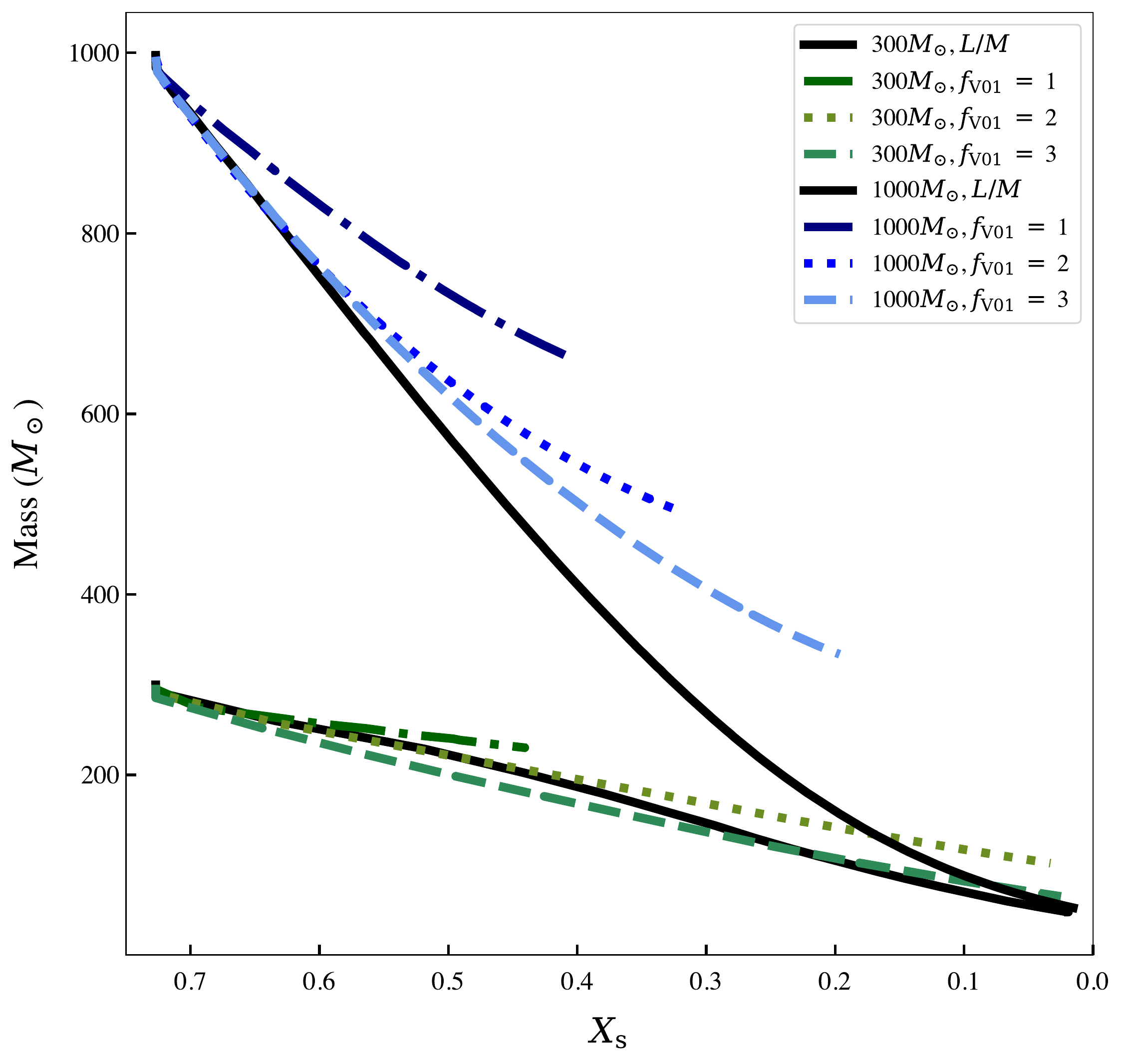}
    \caption{Mass evolution as a function of $X_{\mathrm s}$ for models with $M_\mathrm{init}$ $=$ 300\Mdot\ and 1000\Mdot, for varied mass-loss prescriptions. \corr{Blue and green} lines illustrate models with mass-loss rates which have been adopted from the standard \citet{Vink01} prescription with \corr{multiplicative} factors unity \corr{(dash-dotted lines), increased by a factor of 2 (dotted lines) and increased by a factor of 3 (dashed lines)}. In \corr{black }solid lines, models have been calculated with our updated $L/M$ wind prescription shown in equation \eqref{mdotV11}, from \citet{Sabh22} utilising the $M-$, $L-$dependencies from \citet{vink11} for VMS above the `kink'.}
    \label{fig:V01V11-MH}
\end{figure}

\subsubsection{Qualitative differences in wind dependencies}

Figure \ref{fig:V01V11-MH} provides \corr{a comparison of 300\Mdot\ and 1000\Mdot\ models implementing the updated $L/M$ wind recipe from equation \eqref{mdotV11} and the \cite{Vink01} mass-loss rates }with factors $f_{\rm{V01}}$ multiplied by a linear factor of 2 and 3. 

\begin{figure}
    \includegraphics[width = \columnwidth]{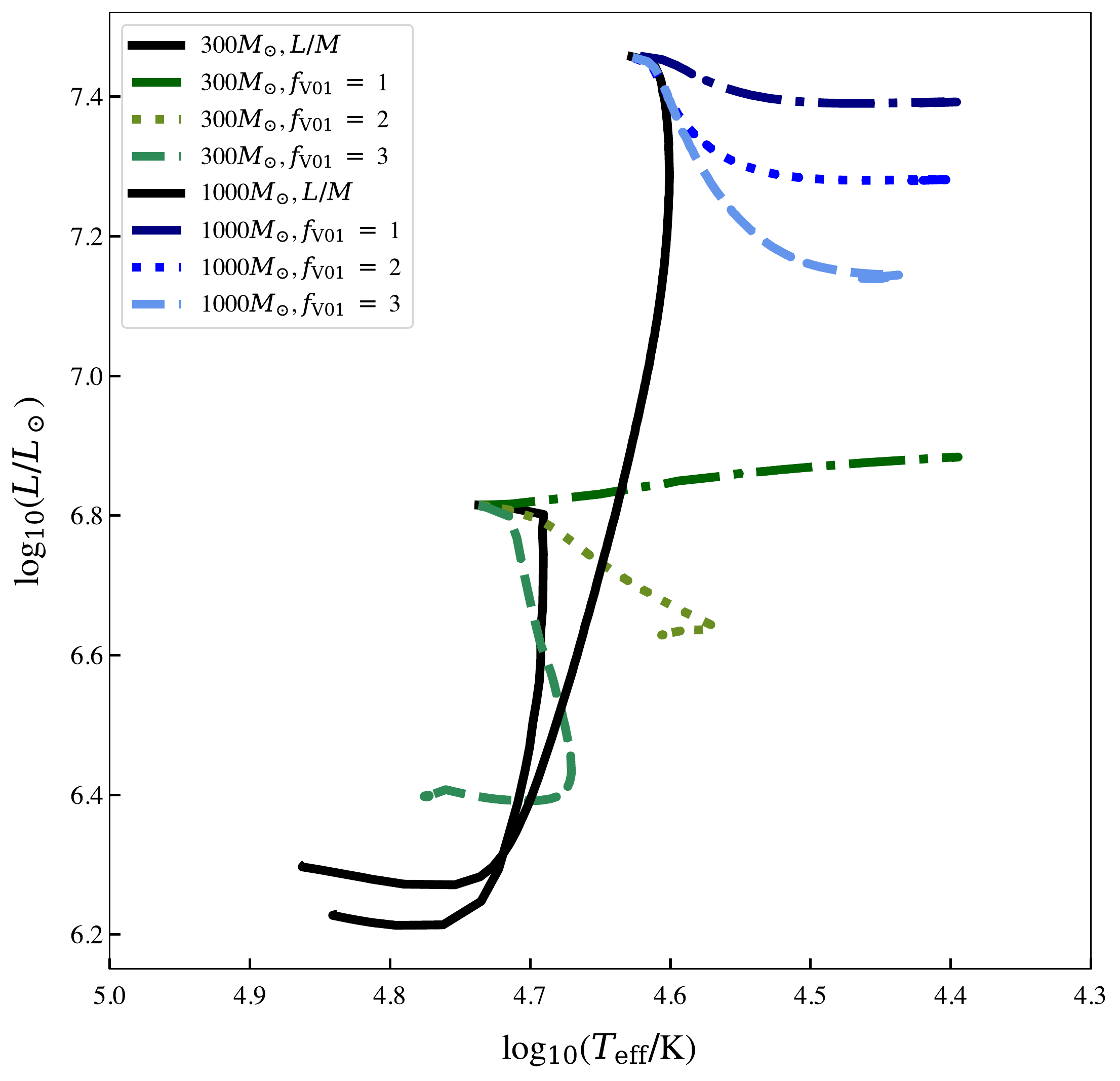}
    \caption{Hertzsprung-Russell diagram showing the evolution of models with $M_\mathrm{init}$ $=$ 300\Mdot\ and 1000\Mdot, for varied mass-loss prescriptions. \corr{Blue and green lines show models with mass-loss rates from the \citet{Vink01} prescription with multiplicative factors $f_{\mathrm{V01}}$ of unity (dash-dotted lines), 2 (dotted lines) and increased by a factor of 3 (dashed lines)}. In \corr{black solid lines, models have been calculated with the $L/M$ wind prescription outlined in equation \eqref{mdotV11}, from \citet{Sabh22}.}}
    \label{fig:V01V11-HRD}
\end{figure}

The stellar models including factors $f_{\rm{V01}}$ of the \cite{Vink01} mass-loss rates \corr{are} shown with factors $f_{\rm{V01}}$ $=$ 1 (dash-dotted), $f_{\rm{V01}}$ $=$ 2 (dotted) and $f_{\rm{V01}}$ $=$ 3 (dashed) for $M_\mathrm{init}$ $=$ 300\Mdot\ (green) and 1000\Mdot\ (blue). We find that arbitrarily increasing the O star rates leads to a wide dispersion of final masses, luminosities and effective temperatures (Fig. \ref{fig:V01V11-HRD}). However, this changes completely when instead employing the VMS recipe from \cite{Sabh22} above the transition point \citep{VG12}. Models including the revised recipe outlined in \cite{Sabh22} with $L/M$-dependent rates are shown for each initial mass (black solid lines). The qualitative change in behaviour is most evident in the final mass dispersion, with models implementing $f_{\rm{V01}}$ $=$ 1 to 3 giving a range of final masses at core H-exhaustion. As evident from the black lines in Fig.\,\ref{fig:V01V11-MH}, we obtain a convergence on the TAMS mass, regardless of $M_\mathrm{init}$, \corr{with the \cite{Sabh22} recipe}. This means that in the last 10\% of core H-burning, it \corr{is} impossible to decipher the $M_\mathrm{init}$ of stars that are now in the range 50-100\Mdot. This is critical when considering observations of WNh stars from \corr{the} VFTS (Table \ref{tab:VFTS}) where \textit{low surface H abundances allow for a wide interpretation of initial masses}, especially since the spread in $T_{\mathrm{eff}}$ is 0.1dex and luminosities are observed up to  $\mathrm{log\corr{_{10}}}\;(L/\rm{L}_\odot)$ $\sim$ 6.6. 

In Fig.\,\ref{fig:V01V11-HRD}, we show the Hertzsprung-Russell diagram (HRD) of the same set of models discussed above, with models evolving from the one ZAMS location for each mass. Despite the different $M_\mathrm{init}$ of 300\Mdot\ and 1000\Mdot, the evolution of the VMS-recipe models (black tracks) is very similar, especially at H exhaustion \corr{($\mathrm{log\corr{_{10}}}\;(L/\rm{L}_\odot)$ $\sim$ 6.2)}. The steep drop in luminosities reflects the high mass-loss rates, leading to a significant drop in mass. In this regard, the VMS tracks look similar to evolutionary tracks of classical WR stars \citep[see, e.g.,][]{Higgins+21}. In contrast, the \cite{Vink01} recipe models evolve redwards towards cooler temperatures, despite also displaying a slight drop in luminosity for the $f_{\mathrm{V01}}$ $=$ 2-3 cases. The track which displays a WR-like behaviour relates to the model with $M_\mathrm{init}$ $=$ 300\Mdot\ and $f_{\mathrm{V01}}$ $=$ 3. For the WNh stars listed in Table \ref{tab:VFTS}, it would be impossible to distinguish between an $M_\mathrm{init}$ of 300\Mdot\ and 1000\Mdot\ in Fig. \ref{fig:V01V11-HRD}. 

\begin{figure}
    \includegraphics[width = \columnwidth]{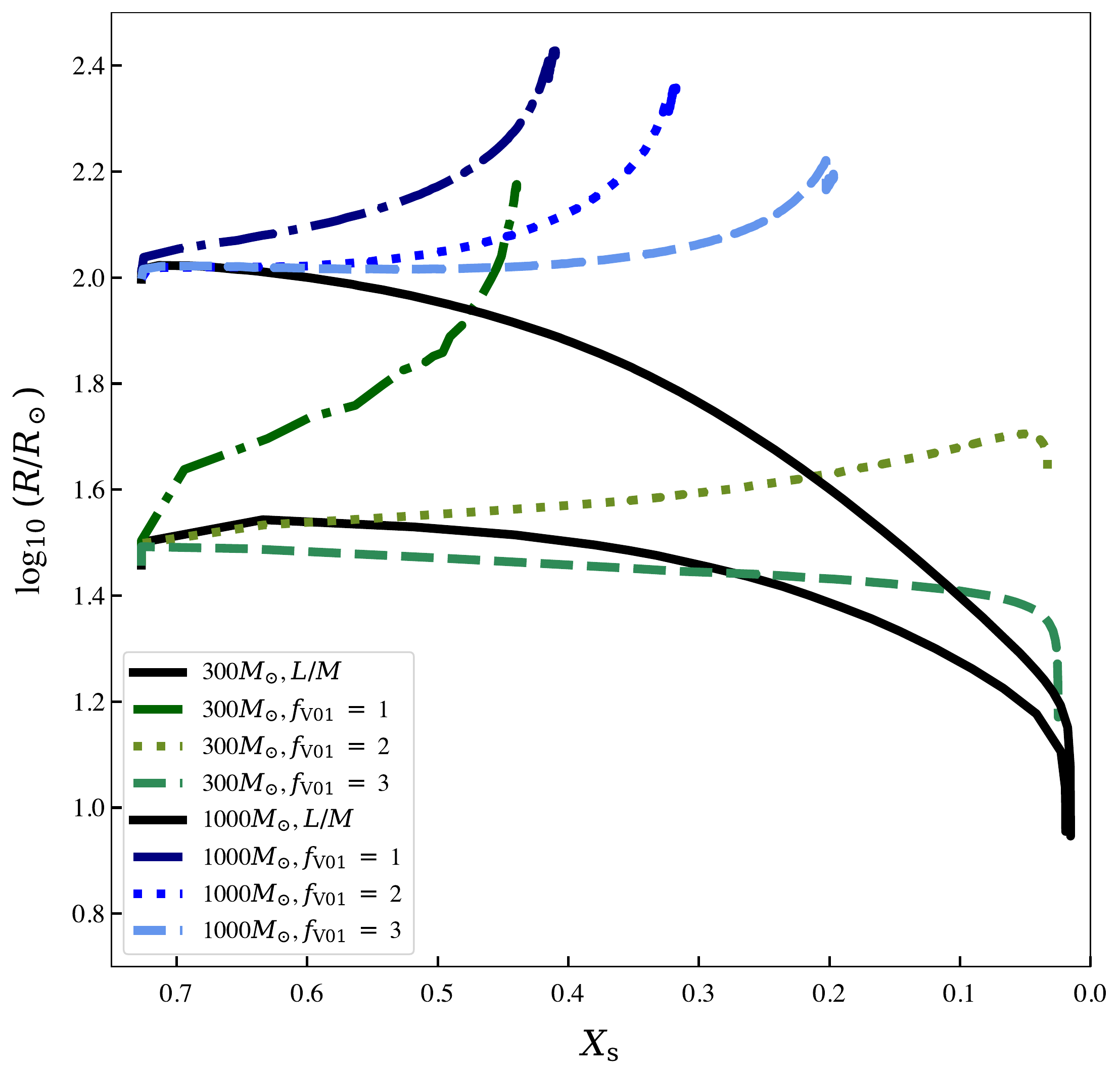}
    \caption{Stellar radius as a function of $X_{\rm s}$ for models with $M_\mathrm{init}$ $=$ 300\Mdot\ and 1000\Mdot\ for varied mass-loss prescriptions\corr{, where colours are as outlined in Fig. \ref{fig:V01V11-MH}.} }
    \label{fig:V01V11-RH}
\end{figure}
The evolution of stellar radii as a function of $X_{\rm s}$ abundance is shown in Fig.\,\ref{fig:V01V11-RH} for $M_\mathrm{init}$ $=$ 300\Mdot\ \corr{($\mathrm{log\corr{_{10}}}\;(R/\rm{R}_\odot)$ $=$ 1.5 at ZAMS) and 1000\Mdot\ ($\mathrm{log\corr{_{10}}}\;(R/\rm{R}_\odot)$ $=$ 2.0 at ZAMS)}, with mass-loss rates as outlined in Fig.\,\ref{fig:V01V11-MH}. The lowest mass-loss rates with $f_{\mathrm{V01}}$ $=$ 1 (dash-dotted) demonstrate that without strong mass loss, VMS inflate due to their proximity to the Eddington limit. With increasing factors ($f_{\mathrm{V01}}$ $=$1-3) of the \cite{Vink01} rates, \corr{the stellar radii become increasingly compact (by up to a factor of five) as the stars lose mass and have lower values of $\Gamma$.} Interestingly, the models which implement \cite{Sabh22} rates (black solid lines) accounting for this proximity to the Eddington limit by increasing $\dot{M}$\, with M and L, \corr{remain} very compact through the MS evolution, with an indistinguishable $\mathrm{log\corr{_{10}}}\;(R/\rm{R}_\odot)$ for both $M_\mathrm{init}$ by the TAMS  \corr{(at $\mathrm{log\corr{_{10}}}\;(R/\rm{R}_\odot)$ $=$ 1.0)}.

These results highlight that while it remains evident that the quantitative mass-loss rates are important for calculating the exact $M_\mathrm{init}$ for a particular VMS, the evolutionary behaviour is drastically changed by a fundamental, qualitative revision of wind dependencies, such as in high $L/M$ situations. The updated wind recipe by \cite{Sabh22} leads to important consequences not only for the mass evolution and stellar radii of VMS but also \corr{the evolution of luminosity and temperature since these stars may evolve at a constant luminosity and towards cooler effective temperatures, or have a steep drop in luminosity while maintaining constant effective temperatures}.  

\subsubsection{Initial mass boundaries}\label{boundaries}
The initial masses of VMS, \corr{predicted from current masses and the estimated mass lost throughout their evolution, is highly dependent on which mass-loss rates are invoked}. While for canonical O stars a factor of $f_{\rm{V01}}$ $<$ 1 has been discussed \citep[e.g.][]{Renzo+2017, bjorklund21}, for stars above $\sim$ 100\Mdot\ mass-loss rates should be enhanced with $f_{\rm{V01}}$  $>$ 1. As a result we compare factors of the \cite{Vink01} O-star mass loss rates ($f_{\rm{V01}}$ $=$ 1,2,3) with rates from \cite{Sabh22}. We find that artificially increasing O-star winds by a factor of \corr{2 to 3} does not alter the qualitative behaviour of the mass-loss history. Instead, an enhanced wind dependence in the form of increased M and L dependencies allows for an appropriate representation of VMS winds. We can estimate the initial mass of a star for a given mass-loss rate. For a factor of the standard \cite{Vink01}, we can integrate \corr{from the initial mass $i$ to the dynamical mass $d$ }over the mass lost at a given age, for example $\sim$ 2\,\corr{Myr}, the estimated cluster age of 30 Dor \citep{Schneider18},\corr{ \begin{equation}\label{integral}
{M_{*}} = {M_{\mathrm{init}}} - \int_{{\mathrm{i}}}^{{\mathrm{d}}} \dot{M} \, {\Delta t}
\end{equation} where }$\Delta$t is estimated for the star's current mass and cluster age. Moreover, this method can be used for a detached eclipsing binary which provides dynamical masses and a constant age for both components \citep{Higgins} leading to \corr{the same, } interchangeable $\Delta$t for the primary and secondary.

Therefore, with the current masses of the primary ($M_{\rm{P}}$) and secondary ($M_{\rm{S}}$), with an estimated initial mass, $M_{\rm{i}}$. The relevant $\dot{M}$ can be calculated and compared to the \cite{Vink01} rates for \ZLMC\, to provide a predicted factor of increase. We provide an example for the components of R144 in equations \eqref{int_P} and \eqref{int_S} for their assumed current $\dot{M}$\, (see Table \ref{tab:R144} for stellar parameters)\corr{, \begin{equation}\label{int_P}
{M_{\mathrm{P}}} ={M_{\mathrm{i}}} - \left[\dot{M}\ \Delta t\right]_{{M_{\mathrm{i}}}}^{74}.
\end{equation}
\begin{equation}\label{int_S}
M_{\mathrm{S}} = M_{\mathrm{i}} - \left[\dot{M}\ \Delta t\right]_{M_{\mathrm{i}}}^{69}.
\end{equation}}
Applying equation \eqref{integral} and integrating for the primary and secondary at an assumed age \corr{($\Delta$t)} allows to simplify the calculation to the following equation for the primary:
\begin{equation}\label{P_int_f}
 \corr{   74\Mdot\ = M_{\mathrm{i}} - \left[(102.6 \Mdot) - (b)\right]}
\end{equation} where b = $\dot{M}_{i}$\ $\times$ $(2 \times 10^{6})$, for an initial mass-loss rate $\dot{M_{i}}$\ which is scaled with a factor $f_{\rm{V01}}$ of the standard \cite{Vink01} rates. 
More generally, we can express the factor $f_{\rm{V01}}$ of the \cite{Vink01} recipe for a given initial mass as: \begin{equation}
  \corr{  f_{\rm{V01}} = \frac{{b}}{\rm{\Delta t}} \times \frac{1}{\dot{M}_{\mathrm{i}}}}.
\end{equation} The calculated factors \corr{of $f_{\rm{V01}}$} of $\dot{M}$, for a range of $M_\mathrm{init}$ $=$ 100-1000\Mdot, lie in the range 1.0-2.6, with $f_{\rm{V01}}$ $=$ 2.59 for the highest $M_{\rm{i}}$. While this solution provides an estimate for the required increased wind rates for VMS up to 1000\Mdot\ reproducing current day masses of the order 50-100\Mdot, the motivation to \corr{arbitrarily increase these base rates by a factor of 1 to 3 of the \cite{Vink01} recipe may not be justified from wind theory or applicability of the \cite{Vink01} prescription in this mass range. Moreover, increasing the mass-loss rate by a factor $f_{\rm{V01}}$ does not account for the realistic wind physics for driving such a strong wind.} Instead, the VMS study by \cite{vink11} demonstrates an increase in mass-loss rate as a function of $L/M$ due to the proximity to the Eddington limit. 

\begin{figure}
    \includegraphics[width = \columnwidth]{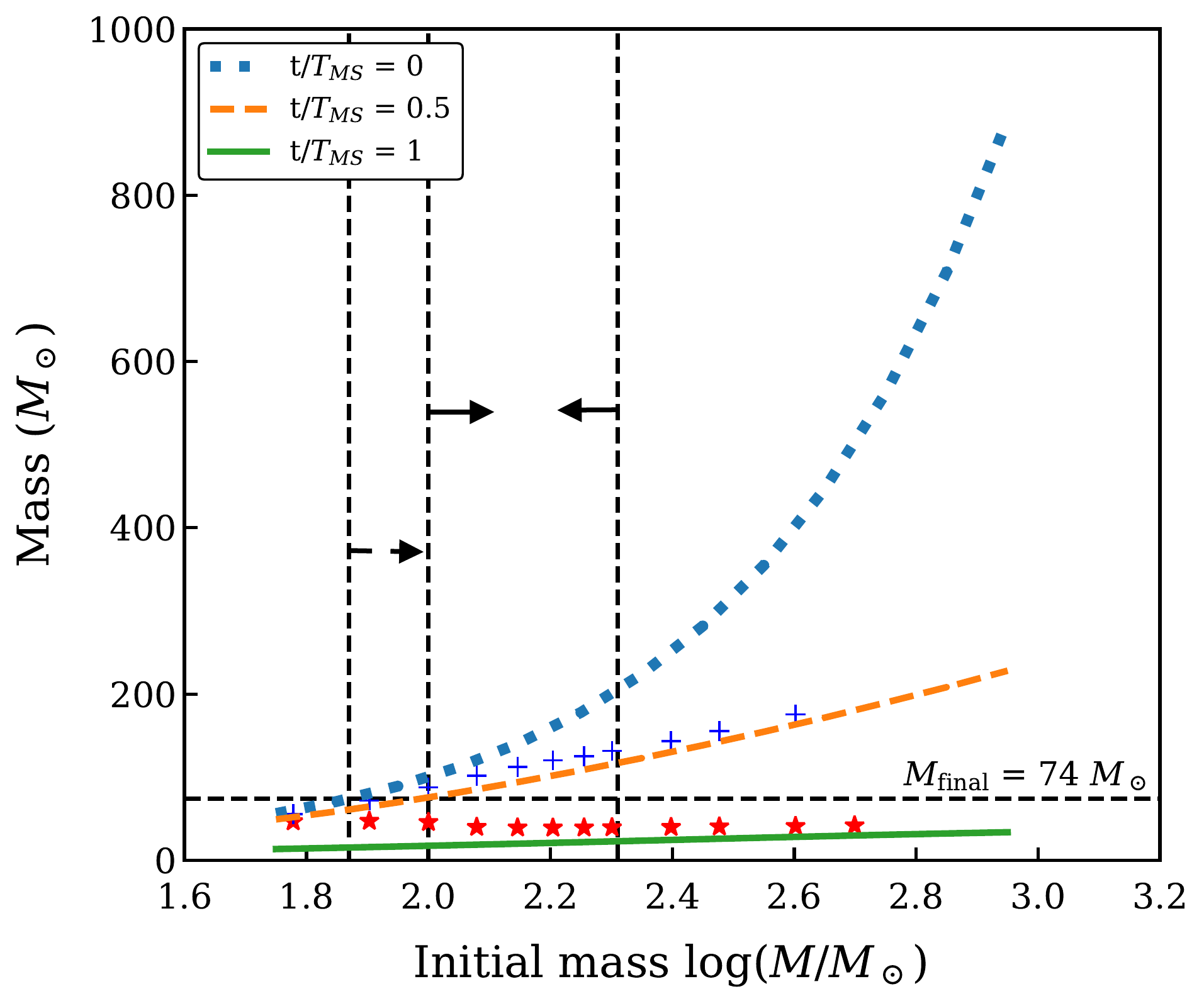}
    \caption{\corr{Stellar} mass as a function of initial mass \corr{for various fractions of the main sequence lifetime, at $X_{\mathrm{c}}$ $=$ 0.7 (ZAMS), 0.35 (HAMS) and 0.0001 (TAMS)}. The ZAMS (dotted blue line), Half-way through MS (HAMS, dashed orange line), and TAMS (solid green line) are shown. These coloured lines represent a \corr{numerically integrated} evolutionary mass calculated iteratively from an initial mass with mass loss calculated from Eq.\,\eqref{mdotV11} and an M-L relation dependent on the \corr{chemical-homogeneous evolution} relation adopted from \citet{Graefener+2011} for VMS. Symbols denote MESA model results for masses at TAMS (red stars) and half-way through the MS (blue pluses). Black dashed lines dictate the upper and lower boundaries for constraining the initial mass range, with solid black arrows highlighting the boundaries. The dashed arrow shows the current mass of the primary component of R144 $M_{\mathrm{dyn}}$ $=$ 74\Mdot.}
    
    
    \label{fig:Gautham_Limits}
\end{figure}

To provide a range of possible $M_\mathrm{init}$ for R144 in the first instance, and for a reduced number of dependencies, we have simplified the \cite{vink11} recipe such that the observed $T_{\rm{eff}}$ and \ZLMC\ have been summed into the base rates, giving a $\dot{M}$ as a function of L and M. For a given M-L relation, we can further reduce the dependencies such that our wind rates are solely a function of mass. We adopt an M-L relationship from \cite{Graefener+2011} relevant for masses 60-4000\Mdot\ with a H mass fraction of 70\% representative of the ZAMS position. The given L-M translation is outlined in equation \ref{MLrelation},
\begin{equation}\label{MLrelation}
    \mathrm{log\corr{_{10}}}\;(L/\rm{L}_\odot) = \;2.12 + 2.39\;\mathrm{log\corr{_{10}}}\; (M/\rm{M}_\odot)\; - 0.20\;\mathrm{log\corr{_{10}}}\;(M/\rm{M}_\odot)^{2}.
\end{equation}
Adopting eq. \eqref{MLrelation} into \cite{vink11} rates from eq.\eqref{mdotV11}, provides a  $\dot{M}$ for a wide range of $M_{\mathrm{init}}$. In this case, we assume an M-L relation for a VMS at ZAMS ($X_\mathrm{c}$ $=$ 0.7) with \ZLMC. The resulting eq.\eqref{ZAMS_H=0.7} as a function of M provides possible solutions of initial mass ranges for a given $\Delta$t, where $\Delta$t is equivalent for the coeval primary and secondary components, 
\begin{equation}\label{ZAMS_H=0.7}
\begin{split}
\mathrm{log\corr{_{10}}} \dot{M}\ (\Mdot\mathrm{yr}^{-1}) = 
-16.48\; + 7.45\;\mathrm{log\corr{_{10}}}(M/\rm{M}_\odot)\;   -0.97\;\mathrm{log\corr{_{10}}}(M/\rm{M}_\odot)^{2}.
\end{split}
\end{equation}
We can calculate theoretical boundaries which limit the possible $M_\mathrm{init}$ range of R144. Figure \ref{fig:Gautham_Limits} illustrates the evolutionary mass as a function of the $M_{\mathrm{init}}$, considering a wide range of possible initial masses $\mathrm{log\corr{_{10}}}$ $\mathrm{(M/\rm{M}_\odot)}$ 1.8-3.0. 
The dashed black lines illustrate boundaries on the possible $M_\mathrm{init}$ due to conditions set by the current dynamical mass of R144 (primary). The first (left) lower limit \corr{at 80\Mdot\ }is set by a lower mass limit where below this line stars would not be able to retain 74\Mdot\ in order to reproduce the dynamical masses. The second (left) boundary \corr{at 100\Mdot\ }dictates that below this limit stars would not be able to maintain 74\Mdot\ in the second half of the MS (see the orange dashed line), and since R144 is close to the TAMS, we can exclude masses below this (middle) boundary line. The upper limit (right black dashed line) \corr{at 200\Mdot\ }represents the maximum $M_\mathrm{init}$ that R144 may have evolved from. 

The higher mass solution consists of fully mixed, chemically homogeneous VMS above $\sim$ 200\Mdot. These stars have surface abundances including H which closely follow their central values. In other words, \textit{the $X_{\rm s}$ abundance is a proxy for the MS age}. Given that the $X_{\rm s}$ abundance of the R144 primary is 0.35, this means \corr{that a H-burning} star above 200\Mdot\ is half-way through their MS. The upper limit can now be obtained by realising that these stars, even with high mass-loss rates, cannot reach the current mass of 74\Mdot (primary) half-way through the MS. This constraint is unique to R144 as we know both the mass and the $X_{\rm s}$ abundance of 0.35. We would not be able to derive any upper limits on the $M_\mathrm{init}$ if the surface abundance \corr{is} lower than $\sim$ 0.25 (see Fig.\ref{fig:MH}).

We also show that breaking the degeneracy of the $M_\mathrm{init}$ via a \corr{numerical} method highlights the crucial importance of the $X_{\rm s}$ abundance on the possible evolutionary history of VMS. For instance, let $M_\mathrm{*}$ be the current mass of a star with a known location in the HRD. Given this information, it is nearly impossible to pinpoint the $M_\mathrm{init}$ of the star as discussed previously. Let $M_\mathrm{init}$ be the initial mass of the star and let the current age of the star be $t_\mathrm{*}$ and $T_\mathrm{MS} (M_\mathrm{init})$ be the main sequence lifetime of the star as a function of its initial mass. \corr{T}his gives the first constraint on the initial mass of the star, that is for a single star with no binary effects, $M_\mathrm{init} > M_\mathrm{*}$. These two quantities can be related using the relation in equation \ref{integral}.

The initial mass of the star can either be (1) greater than $\sim 200 \; \rm{M}_\odot$ (extreme) resulting in CHE throughout the \corr{main sequence lifetime $T_{\rm{MS}}$}, or (2) less than $\sim 200 \; \rm{M}_\odot$ (moderate) and have either homogeneous evolution close to H exhaustion or no homogeneous evolution at all. Regardless one can relate the surface H abundance ($X_{\rm s}$) to the core abundance ($X_\mathrm{c}$), \corr{with an initial $X_\mathrm{c}$ of 0.7,} as follows,
\begin{equation}
\begin{split}
X_{\rm S}\corr{ \gtrsim } X_\mathrm{c} = -0.7 \dfrac{t}{T_\mathrm{MS}} + 0.7.
\end{split}
\label{eq:X_s_X_c}
\end{equation}
The equality $X_{\rm S} \approx X_\mathrm{c}$ holds true for $M_\mathrm{init} \gtrsim 200 \; \rm{M}_\odot$, while the inequality holds true for $M_\mathrm{init} \lesssim 200 \; \rm{M}_\odot$. The second equality is based on the assumption that the star burns H at a constant rate during its MS.

For a known $X_{\rm s}$, Eq. \eqref{eq:X_s_X_c} provide\corr{s} some constraints on the age of the star compared to its MS lifetime. For example, \corr{an }$X_{\rm s}$ of 0.35 implies the star is either half way through the MS (if fully homogeneous) or in the second half of its MS. With the current $X_{\rm s}$, the current mass of the star and constraints on its age, one can obtain lower and upper limits for the $M_\mathrm{init}$\ of the star. The lower limit is based on the star being able to reach the current mass only at ages excluded by the constraint defined in Eq. \eqref{eq:frac_t_Tms},
\begin{equation}
\begin{split}
\dfrac{t_\mathrm{*}}{T_\mathrm{MS}} \geq 1 - \dfrac{X_{\rm S}}{0.7}.
\end{split}
\label{eq:frac_t_Tms}
\end{equation}
The upper limit usually concerns the CHE where the equality in Eq. \eqref{eq:frac_t_Tms} holds true, and the star is unable to reach the current mass within the time $t_\mathrm{*}$. While these limits exclude a substantial part of the initial mass parameter space, more accurate constraints can be obtained by detailed stellar evolution models. 

\begin{figure}
    \includegraphics[width = \columnwidth]{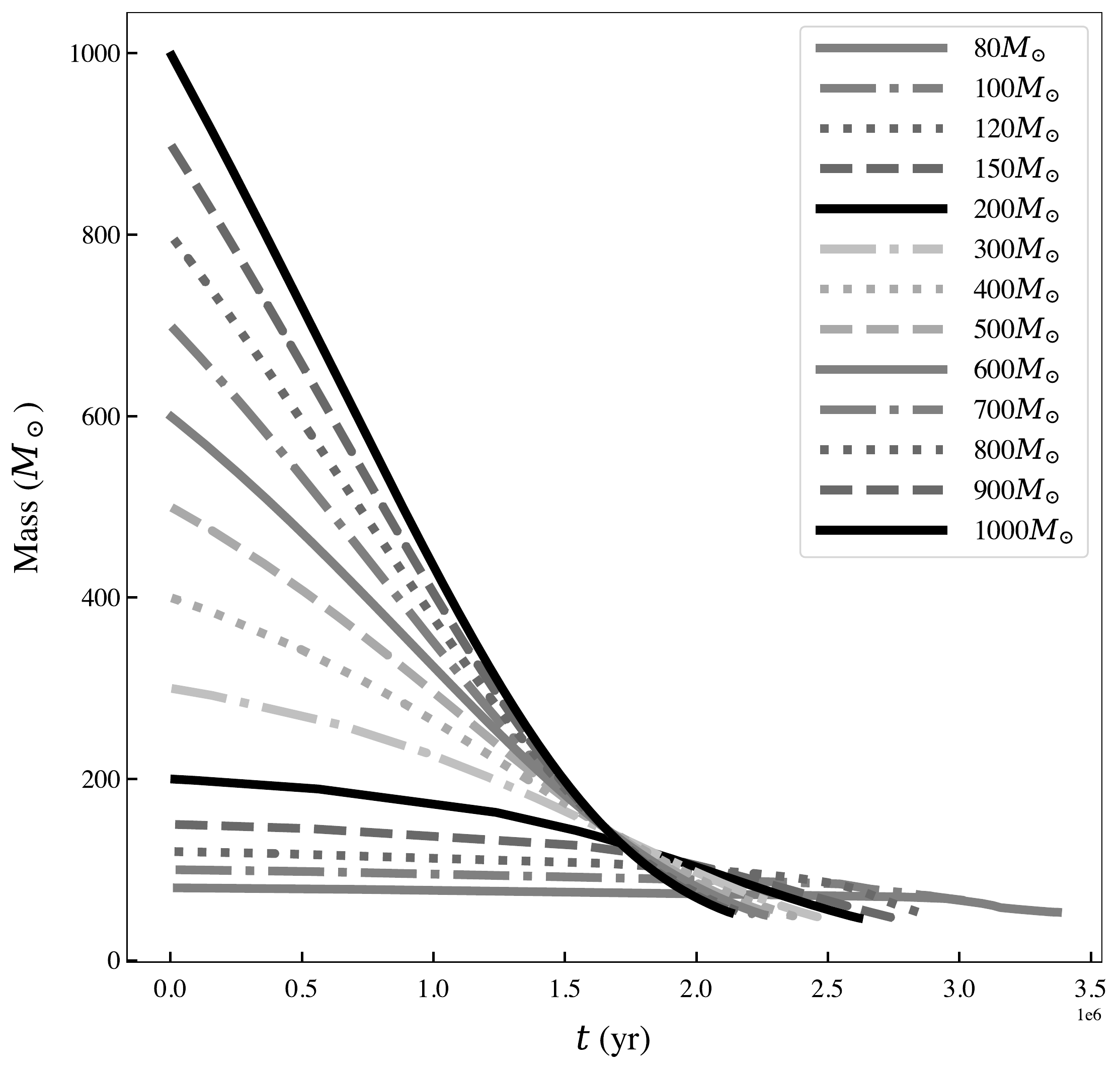}
    \caption{Mass evolution of our standard grid of models with masses 80-1000\Mdot\ evolving from ZAMS \corr{to} TAMS. Alternating \corr{grey lines represent differing masses in the} grid. }
    \label{fig:Mt_all}
\end{figure}
\begin{figure}
    \includegraphics[width = \columnwidth]{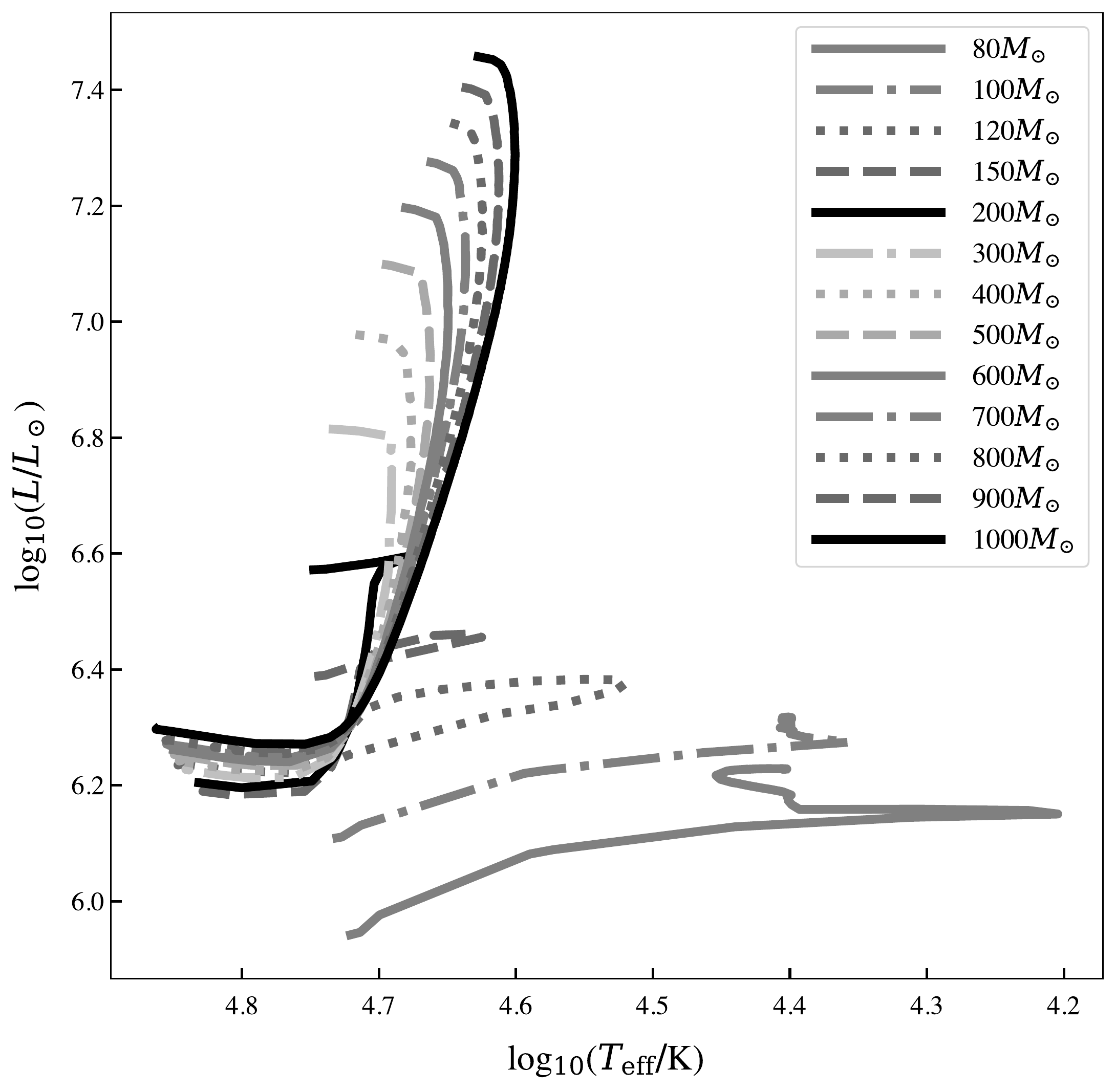}
    \caption{Hertzsprung-Russell diagram of the evolution of our standard grid of models with masses 80-1000\Mdot\ evolving from ZAMS \corr{to} TAMS. Alternating \corr{grey lines} represent change in mass through the grid.}
    \label{fig:HRD_all}
\end{figure}

\subsection{Stellar models}\label{modelsSect}
In the remaining sections, we compare with models which have \corr{adopted} the wind prescription by \citet{Sabh22} which is updated for VMS with an $L/M$ dependency as outlined in \citet{vink11}. This is incorporated beyond the observed transitional mass-loss rate \citep[log$_{10}$ $\dot{M}$ $=$ -5.0,][]{VG12, Sabh22}, leading to a rough transition mass \corr{(121.8\Mdot)} and luminosity \corr{($\mathrm{log_{10}}\;(L/\mathrm{L}_\odot)$ $=$ 6.31)} for the LMC. The mass-loss routine adopts \corr{exponent}s for L and M as in \citet{vink11}, shown in equation \ref{mdotV11}.

\subsubsection{Mass turnover}
Figure \ref{fig:Mt_all} provides an overview of the mass evolution with stellar age for all models in our standard grid \corr{from 80-1000\Mdot} until H exhaustion. Moderate mass ($M_\mathrm{init}$ $<$200\Mdot) models demonstrate a shallow gradient in the mass, due to the lower $\dot{M}$\ rates below the `kink', active below the transition point. Models with extreme $M_\mathrm{init}$ above $\sim$200\Mdot\ display a much steeper drop in mass and converge to a \corr{mass of 140 $\pm$10\Mdot\ at 1.675 $\pm$0.25 Myr}. At this point, it is impossible to determine the $M_{\mathrm{init}}$. Moreover, the evolution of both moderate and extreme mass ranges, overlap beyond $\sim$ 1.6 \,\corr{Myr} leading to a degeneracy in the $M_\mathrm{init}$ whereby the current model may have evolved from any of the models, with the entire $M_\mathrm{init}$ range 80-1000\Mdot\ as a possibility. This is particularly interesting for observations of massive stars in the mass range 50-100\Mdot\ with current ages 1.5-2.5\,\corr{Myr}. \cite{Brands22} also confirms the cluster age of the most massive stars in the R136 cluster of the LMC to be 1-2.5 \,\corr{Myr}, enclosing the turnover mass of our VMS models. This highlights the uncertainty in the possible initial upper stellar mass.

The convergence of the upper initial mass was suggested by \corr{ \cite[][see their fig. 3]{vink18} }where their integrated mass-loss rates, calculated for a constant effective temperature, demonstrated that stars should converge to a mass of $\sim$ 200\Mdot\ independent of $M_\mathrm{init}$ at 2\,\corr{Myr}. These results show that while the quantitative upper initial mass may depend on factors such as $Z$, the qualitative behaviour will depend on the adopted mass-loss \corr{exponent}s. In this case, it is evident that enhanced winds are most appropriate for VMS compared with standard O star dependencies from \cite{Vink01}. Moreover, \cite{Belkus07} and \cite{Yung08} demonstrated that \Zdot\ models with $M_\mathrm{init}$ up to 1000\Mdot\ also showed a mass convergence around 2\,\corr{Myr}, although their mass-loss prescriptions were ad hoc \corr{and} not well physically-motivated. 

In this work, we find a mass convergence at \corr{140 $\pm$10\Mdot, and t $=$ 1.675 $\pm$0.25}\,\corr{Myr}, but more interestingly, we find a mass turnover where the highest mass stars actually produce the lowest \corr{($\sim$50\Mdot)} masses at $\sim$ 1.6$-$2\,\corr{Myr}. The inclusion of equation \ref{mdotV11} for the enhanced mass-loss rates in our models demonstrates that increased mass loss above the `kink' leads to qualitatively different behaviour than for factors of the standard \cite{Vink01} recipe. Fig. \ref{fig:Mt_all} nicely illustrates the mass turnover in the  gradient at $\sim$ 1.6 \,\corr{Myr}, suggesting that for observed masses up to 100\Mdot\ the mass\corr{-loss} history is indistinguishable. 
\begin{figure}
    \includegraphics[width = \columnwidth]{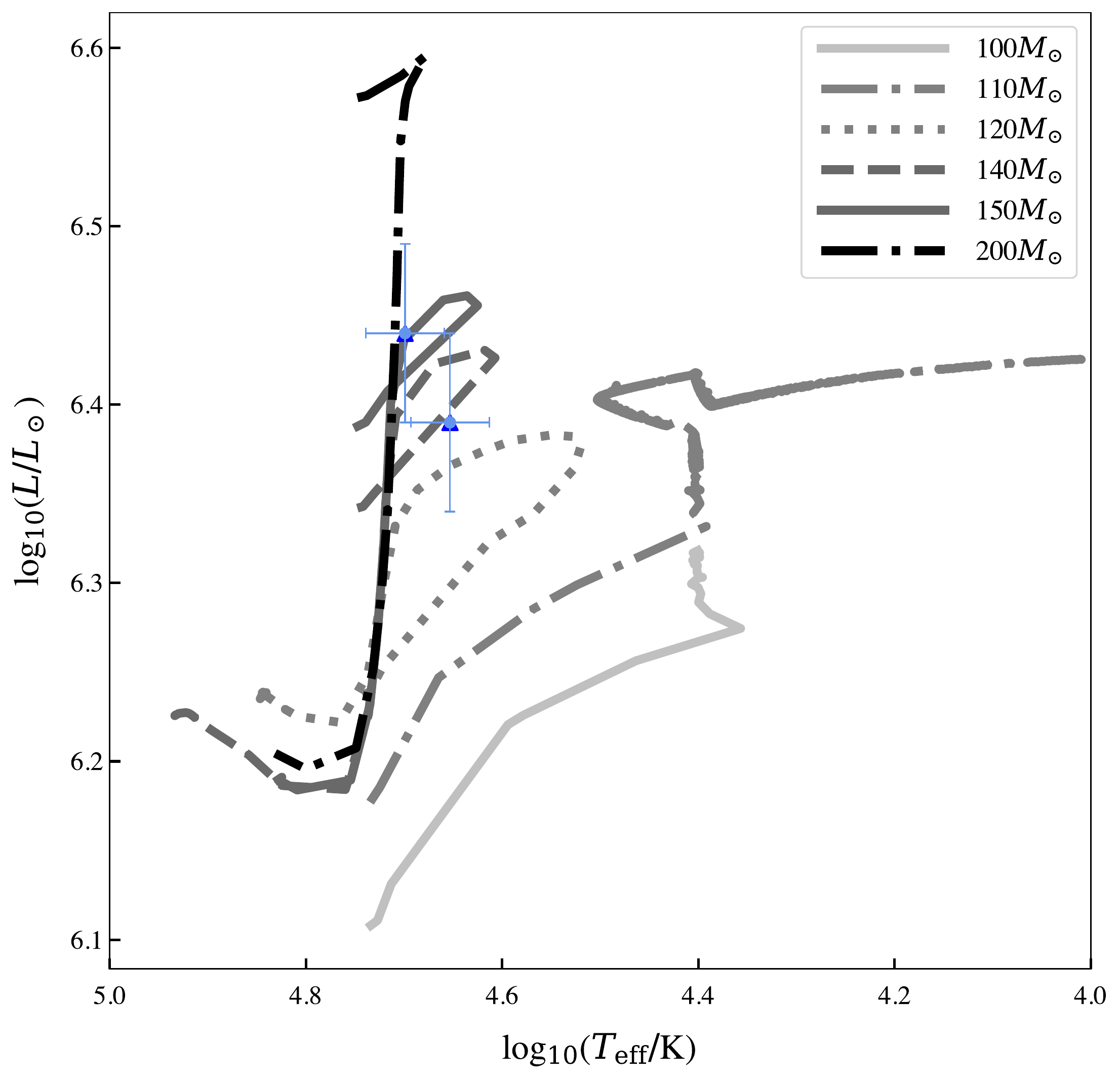}
    \caption{Hertzsprung-Russell diagram of the evolution of our standard models with masses 100, 110, 120, 140, 150, and 200\Mdot\ \corr{(grey scale) evolving from ZAMS to TAMS. Blue} triangles illustrate the primary and secondary components of R144.}
    \label{fig:R144_HRD}
\end{figure}
\begin{figure}
    \includegraphics[width = \columnwidth]{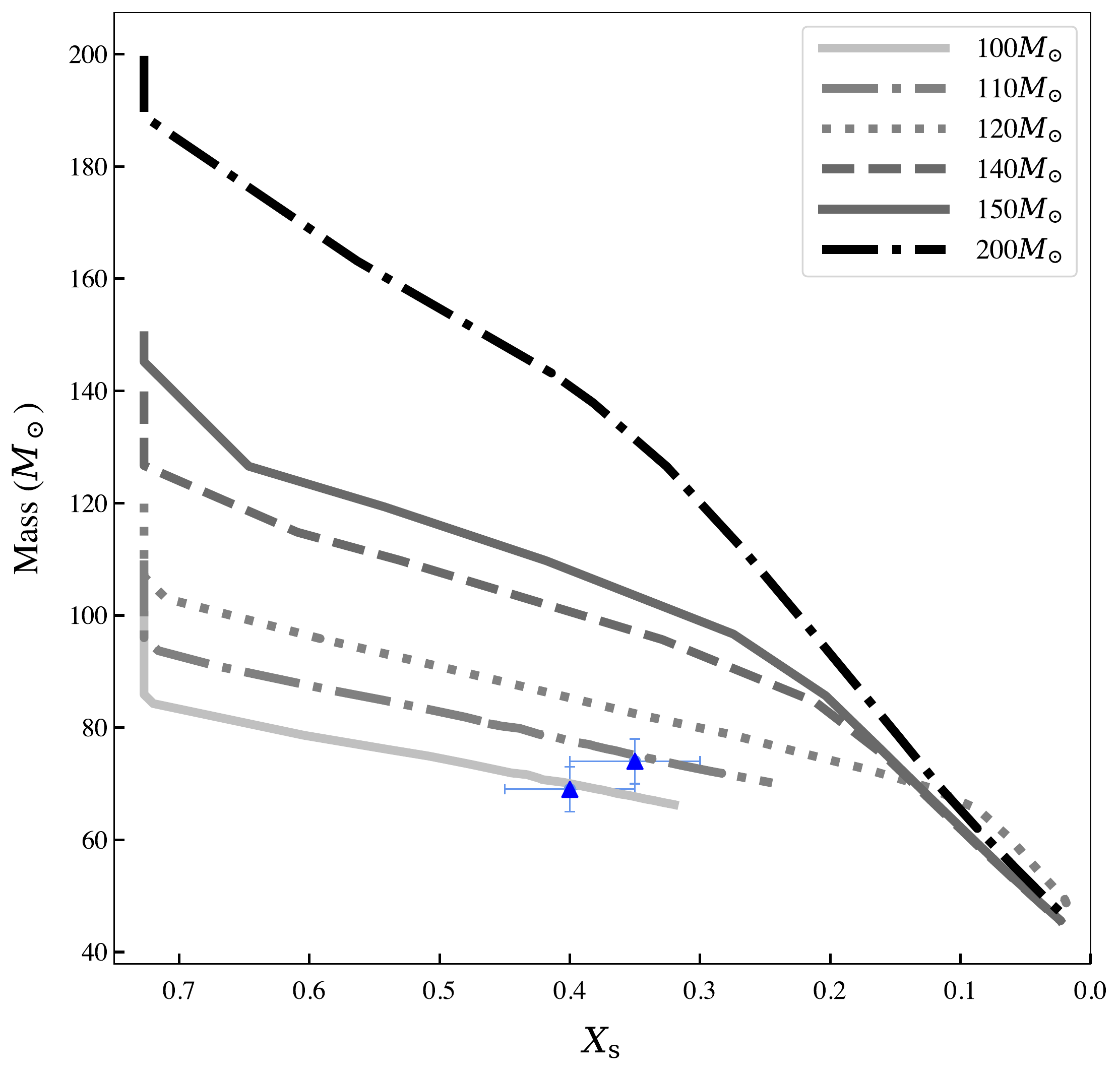}
    \caption{Mass evolution of models \corr{with masses 100, 110, 120, 140, 150, and 200\Mdot\ as a function of $X_{\rm s}$ abundance. Blue} triangles show the primary and secondary components of R144.}
    \label{fig:R144_M_H}
\end{figure}

\subsubsection{Luminosity and effective temperature evolution}
Figure \ref{fig:HRD_all} represents the evolution of our model grid in a Hertzsprung-Russell diagram (HRD). We find that our models with $M_\mathrm{init}$ above 200\Mdot\ have a wide range in luminosity and small variation in effective temperature. This is due to the strong $L/M$-dependence of the wind prescription leading to a significant drop in mass and subsequently luminosity during the MS evolution. While the ZAMS for models in the extreme mass range 300-1000\Mdot\ may differ, ranging up to log$_{10}$ (L/\Ldot) $=$ 7.2 for the 1000\Mdot\ model, the almost linear decline in luminosity leads to indistinguishable evolutionary characteristics in the HRD, with final masses of $\sim$ 50\Mdot\ also leading to similar luminosities and temperatures at the TAMS for this initial mass range. 

On the other hand, lower mass models with $M_\mathrm{init}$ below 200\Mdot, mainly the models with 80-120\Mdot, have an evolutionary behaviour similar to standard O-stars, evolving to cooler effective temperatures during the MS evolution. The lower mass models in this study (80-120\Mdot) are still very massive compared to the more numerous O star populations of e.g. 20-60\Mdot\ and as such experience inflation effects. 

The moderate mass range of 120-200\Mdot\ have characteristics of both subsets, where the initial MS evolution is dominated by an increase in luminosity and a small decline in effective temperature. However, by midway through the MS, these models also turn to hotter temperatures like their more massive (300\Mdot\ and above) counterparts, in a WR-like fashion due to the self-regulatory effect of the $L/M$-dependent winds. From the HRD, we can see that a small uncertainty in $T_{\mathrm{eff}}$ can lead to large uncertainties in the HRD with respect to evolutionary history and initial mass range.

The indistinguishable evolutionary tracks of all VMS models in Fig.\,\ref{fig:HRD_all} ($M_{\mathrm{init}}$ > 200\Mdot) further confirm the degeneracy of the upper initial mass since observations in the effective temperature range of log$_{\rm{10}}$ ($T_{\mathrm{eff}}$/K) $\sim$ 4.7 would have a wide range of possible initial masses, even for a broad range of luminosities due to the nature of the steep drop in mass for all models. The narrow temperature range of VMS models with masses above 200\Mdot\ is in good agreement with the observed temperature range of WNh stars in the 30 Dor cluster \citep[log$_{\rm{10}}$ ($T_{\mathrm{eff}}$/K) $=$ 4.6-4.7 (K),][]{Sabh22}. 

\subsubsection{Reproducing R144}\label{R144sect}
In this work we utilised the most luminous and massive detached eclipsing binary from the Tarantula Massive Binary Monitoring (TMBM) sample. R144 comprises of massive 74\Mdot\ and 69\Mdot\ components with well constrained dynamical masses, see other stellar parameters included in Table \ref{tab:R144}. These dynamical masses and surface abundances provide a useful calibration point for our models of VMS. We tested multiple sets of model inputs covering a range of rotation rates, and core overshooting values. We find that there is a conflict in estimating which initial mass range\corr{, moderate ($<$200\Mdot) or extreme ($>$200\Mdot), }these objects may have originated from when comparing to stellar observables such as luminosities, radii and $X_{\rm s}$ abundances. From Fig.\,\ref{fig:HRD_all} we note that models with $M_{\rm{init}}$ $\geq$200\Mdot\ follow the same evolutionary behaviour in the HRD. Figure \ref{fig:R144_HRD} then highlights that our extreme models with $M_{\rm{init}}$ $\geq$ 200\Mdot\ can reproduce the HRD positions of R144 components, as seen by the overlap of the 200\Mdot\ track with the blue triangles. As mentioned above, the evolution of VMS is dominated by strong mass loss and as a result leads to almost linear evolution in the HRD with a steep drop in luminosities making the initial mass range indistinguishable. From Fig.\, \ref{fig:R144_HRD} we can see that with a moderate uncertainty in the observed luminosities ($\sim$ 0.2 dex) our predictions of $M_\mathrm{init}$ could change from the highest mass range ($>$200\Mdot) to the moderate mass range ($\sim$100\Mdot). The overlap between moderate models (100-200\Mdot) and the highest mass models which evolve chemically homogeneously, correlates with the HRD positions of both components of R144, leading to a wide degeneracy in the possible initial mass range.

In order to break the degeneracy we must test if they are able to simultaneously reproduce the dynamical masses of the system alongside their observed H abundances. We can see from figure \ref{fig:R144_M_H} that our lowered $M_\mathrm{init}$ models of 110\Mdot\ and 100\Mdot\ can reproduce the $M_{\mathrm{dyn}}$ and $X_{\rm s}$. The initial decrease in mass can be noticed for all models in the $M_\mathrm{init}$ range 100-120\Mdot\ at $X_{\rm s}$ $=$ 0.7, since the envelope has not been sufficiently stripped to expose the core H evolution. Similarly, since stars below $\sim$200\Mdot\ do not evolve chemically homogeneously, the $X_{\rm s}$ abundance cannot be used as a direct comparison to the core evolution. 

\begin{figure}
    \includegraphics[width = \columnwidth]{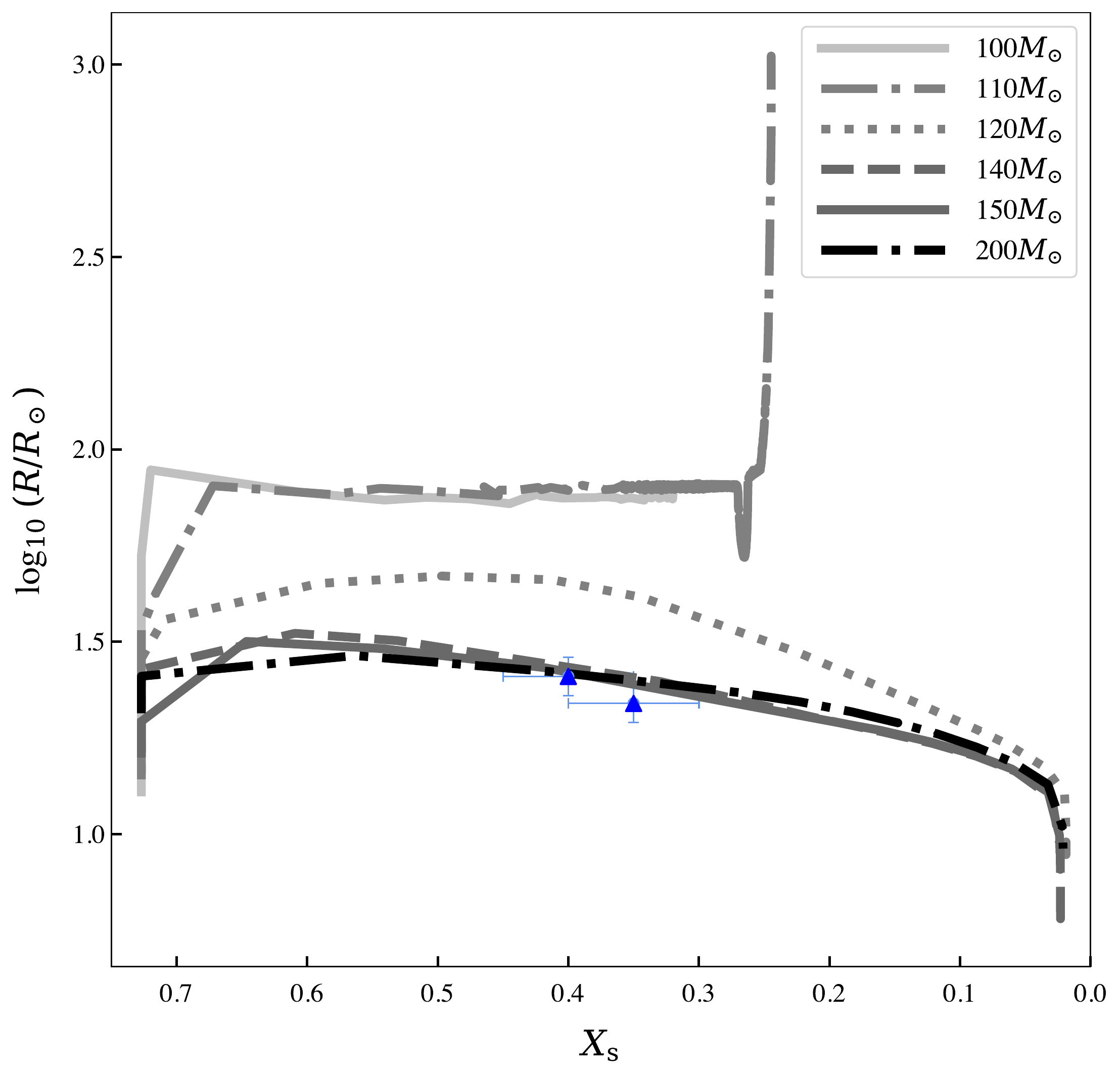}
    \caption{Radius evolution of models as a function of $X_{\rm s}$ from our standard grid with initial masses 100, 110 and 120\Mdot\ in red scale, higher initial masses of 140, 150 and 200\Mdot\ are shown in blue scale. The effects of inflation can be seen in the lowest mass model, whereas the effects of increased mass loss can be seen in the highest mass model. The black triangles depict the primary and secondary components of R144.}
    \label{fig:logRt_R144_R-uncertainty}
\end{figure}

In order to distinguish between \corr{characteristics of VMS evolution} and lower $M_{\rm{init}}$, we tested the effects of mass loss and mixing on observed stellar parameters. We compared the observed stellar radii and $X_{\rm s}$ abundances of the components of R144 with our sets of stellar models in the following section. Fig. \ref{fig:logRt_R144_R-uncertainty} highlights the change in $\mathrm{log\corr{_{10}}}\;(R/\rm{R}_\odot)$ for models with $M_\mathrm{init}$ $=$ 100\Mdot\ and 200\Mdot. Due to the inflation of the 100\Mdot\ model, the stellar radius increases dramatically. However, the more massive model with an $M_\mathrm{init}$ $=$ 200\Mdot\ remains more compact due to the increased effect of stellar winds. We find that the compact radii ($\mathrm{log\corr{_{10}}}\;(R/\rm{R}_\odot)$ $\sim$ 1.4) are only reproduced by higher mass models $\sim$140-200\Mdot. The relatively small inflation of R144 components found by \cite{Shenar21} would suggest that there is a way to have classical Eddington parameters of $\Gamma_{\rm{e}}$ $\sim$ 0.78 while remaining compact enough with a radius of $\sim$ 25\Rdot.

The observed $X_{\rm s}$ abundances of R144 can be used as an important tool for breaking the degeneracy in possible initial mass ranges as well as providing an age proxy. Due to the chemically-mixed nature of stars \corr{more massive than} 200\Mdot, alongside the enlarged ratio of core to total masses of VMS, the $X_{\rm s}$ abundance is a very good indicator of the MS burning stage of evolution. We find that with 35 and 40\% surface H abundances, the primary and secondary are likely products of moderate masses $M_{\rm{init}}$ $\sim$ 100-120\Mdot\ as models with extreme $M_\mathrm{init}$ $=$ 300-1000\Mdot\ only lose enough mass to reach $M$ $<$100\Mdot\ at $X_{\rm s}$ abundances of 20\% and less. 

\begin{figure}
    \includegraphics[width = \columnwidth]{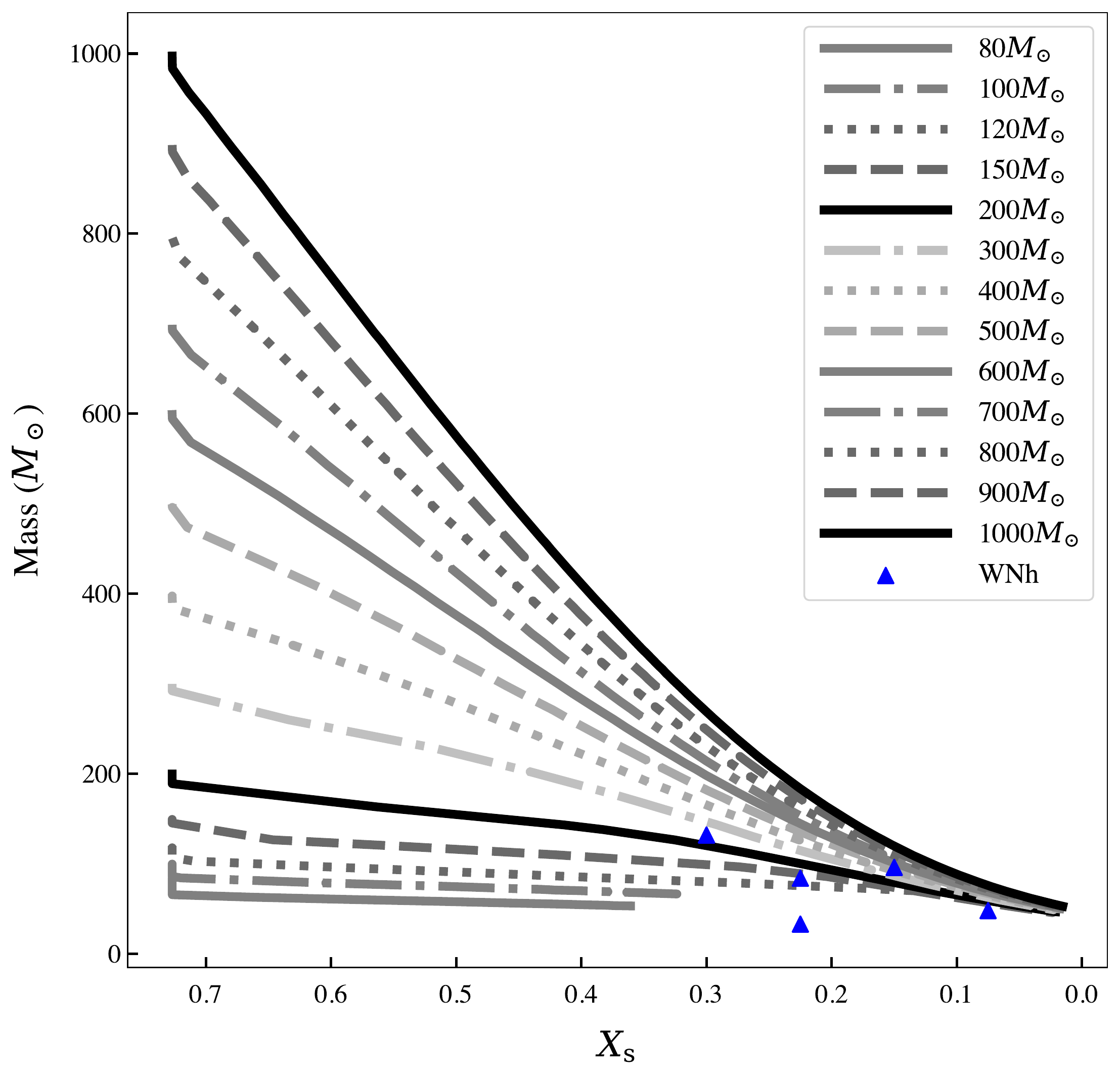}
    \caption{Mass evolution as a function of $X_{\mathrm s}$ for our standard grid of non-rotating models. Our models range from 80\Mdot\ to 1000\Mdot, calculated from ZAMS until core H exhaustion, with alternating grey lines for different initial masses. Blue triangles locate the WNh sample from \citet{best14}.}
    \label{fig:MH}
\end{figure}

\subsubsection{WNh comparison}

Since VMS models above 200\Mdot\ lose mass throughout the MS, and -- crucially due to their chemically homogeneous nature -- also linearly drop in $X_{\rm s}$ during the MS \corr{(see Fig.\,\ref{fig:Xsurf-Xcore})}, the $X_{\rm s}$ can be a probe for the $M_\mathrm{init}$ where the $X_{\rm c}$ is a function of observed current $X_{\rm s}$. Whereas, for moderate mass models (80-200\Mdot) most of the MS is spent with 70\% $X_{\rm s}$ and only decreases towards the end of the MS. This means that while the mass turnover at 1.6\,\corr{Myr} suggests the $M_\mathrm{init}$ is indistinguishable, it can be deciphered if the current mass is compared with the current $X_{\rm s}$ abundance. For instance, a star observed to have 50\% $X_{\rm s}$ remaining but a current mass of 200\Mdot\ could not have formed from a star with an $M_\mathrm{init}$ greater than $\sim$ 300\Mdot\ because sufficient mass would not have been lost by such a VMS during the early stages of the MS. 

\begin{figure}
    \includegraphics[width = \columnwidth]{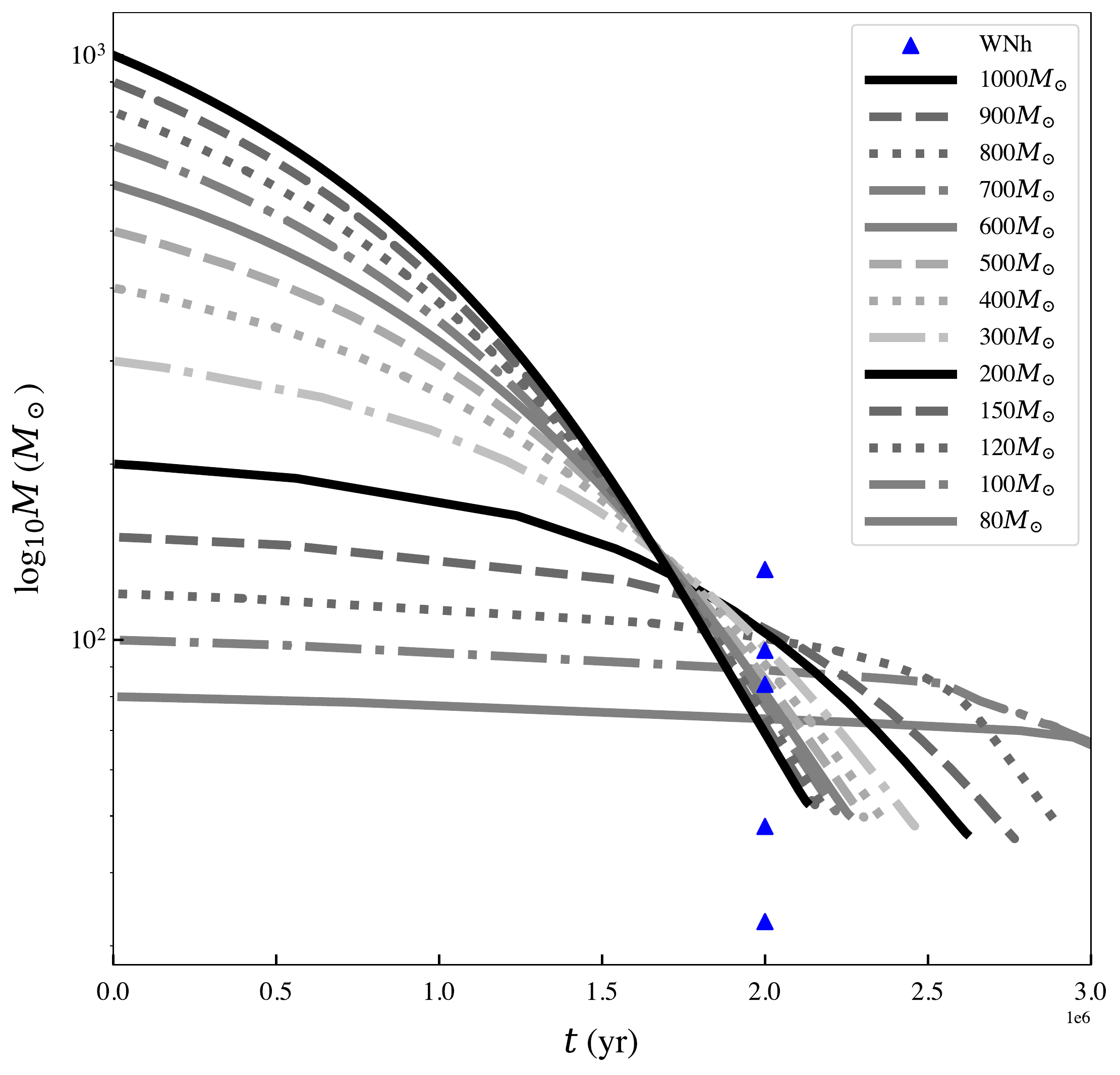}
    \caption{Mass evolution \corr{in logarithmic scale as a function of MS age for models as outlined in Fig.\,\ref{fig:MH}. Blue} triangles locate the WNh sample from \citet{best14}.}
    \label{fig:WNh_age}
\end{figure}

Figure \ref{fig:MH} demonstrates the importance of $X_{\rm s}$ abundances on breaking the degeneracy of possible initial mass ranges. Th\corr{e} sample of WNh stars from \cite{best14} illustrated by blue triangles in Fig.\,\ref{fig:MH} showcase that with \corr{an }$X_{\rm s}$ abundance lower than $\sim$ 30\%, it becomes increasingly difficult to decipher the possible initial mass range for objects which now have a mass of 50-100\Mdot. For instance, at $X_{\rm s}$ $=$ 0.15 and $M_{\rm{*}}$ $=$ 96\Mdot\ the evolutionary tracks in Fig.\,\ref{fig:MH} overlap such that the $M_\mathrm{init}$ of this object is indistinguishable. This means that for VFTS 695 and VFTS 427 (with $X_{\rm s}$ $<$ 0.2), for instance, the $M_\mathrm{init}$ is un\corr{known}. Figure \ref{fig:MH} showcases that when the $X_{\rm s}$ abundance lies below $\leq$ 20\%, a star with an observed mass of 50-100\Mdot\ could originate from a VMS with $M_{\rm{init}}$ $\geq$ 300\Mdot. Moreover, while the age of these WNh stars may be uncertain, \cite{Brands22} finds that the age of the R136 cluster within 30Dor is 1-2.5\,\corr{Myr}. Hence, Fig.\,\ref{fig:WNh_age} further highlights that beyond the turnover mass of 140\Mdot\, at 1.6\,\corr{Myr}, the $M_\mathrm{init}$ of stars observed in the mass range 50-100\Mdot\, are unconstrained. In fact, the most massive stars from our model grid produce the lowest masses \corr{ of $\sim$50\Mdot\ }by $\sim$ 2\,\corr{Myr}. 

\begin{figure}
    \includegraphics[width = \columnwidth]{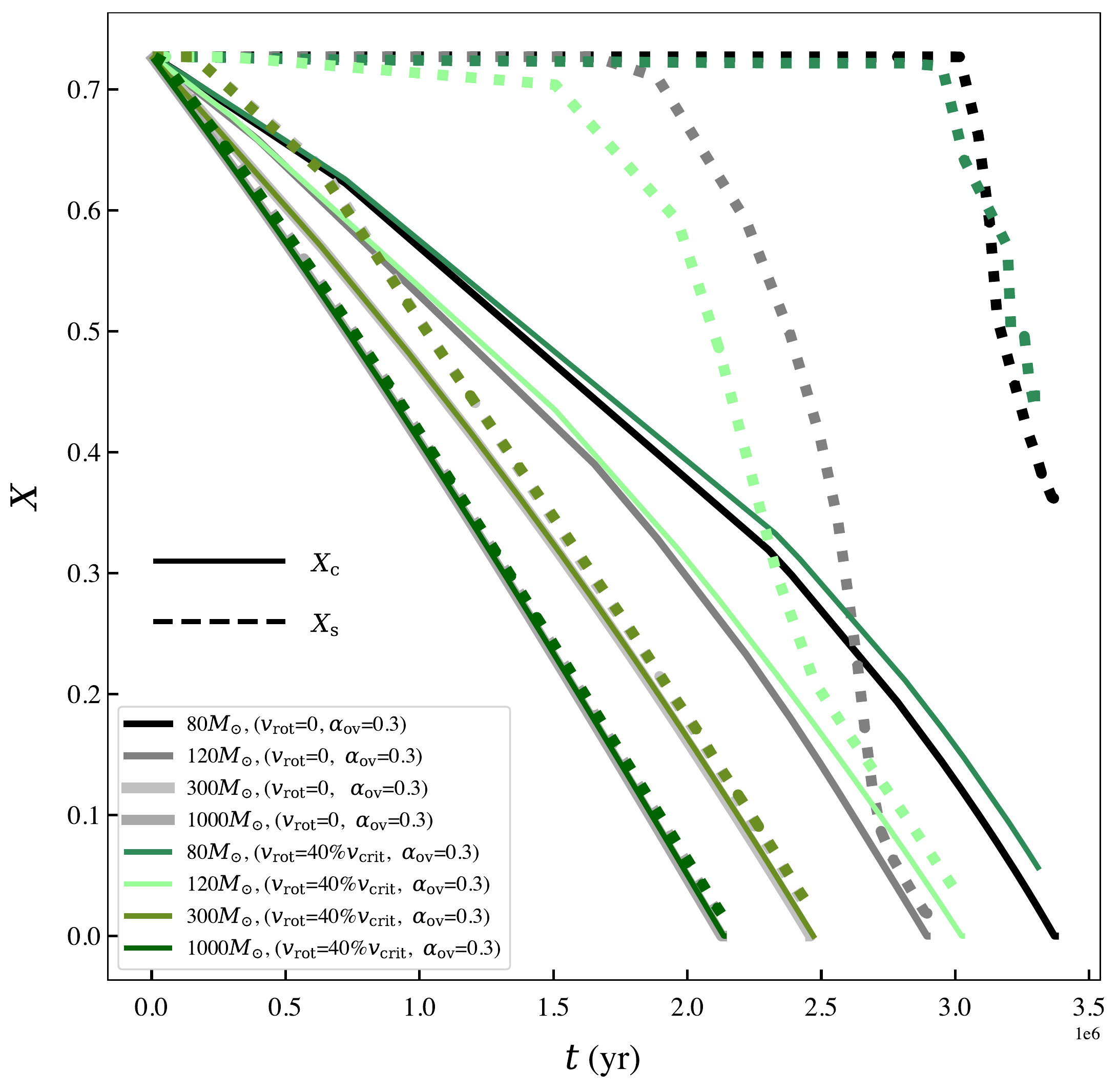}
    \caption{Core $X_{\rm c}$ and surface $X_{\rm s}$ H abundance (solid and dashed lines respectively) as a function of MS age $t$ for stellar models with initial masses 80,\corr{ 120, 300 and 1000\Mdot. We include models with standard inputs in black and grey (non-rotating with \aov\,$=$\,0.3) and models which have 40\% critical rotation in shades of green for different initial masses}.}
    \label{fig:XH}
\end{figure}

Figure \ref{fig:XH} illustrates the evolution of the $X_{\rm s}$ abundance (dashed lines) compared to the $X_{\rm c}$ (solid lines) for models with varied internal mixing. \corr{Models in the lower mass range (80-120\Mdot)} account for different efficiencies in chemical mixing (non-rotating, 40\%\, critical rotation and \aov\, $=$ 0.5) leading to steeper progression of the $X_{\rm s}$ depletion, the VMS models (300\Mdot\ and 1000\Mdot) show that all $X_{\rm s}$ abundances line up exactly with the central H abundances. This means that while a star below $\sim$150\Mdot\ may be able to retain some $X_{\rm s}$\, as the star is not fully mixed and reproduce an observed abundance of 35-40\% at various points during the MS, the more massive models show that a star with $M_\mathrm{init}$ $=$ 300-1000\Mdot\, will only be observed to have 35\% $X_{\rm s}$ at exactly half-way through the MS. In this case the H abundances of 35-40\% provide an additional constraint on the origin of R144. A lowered abundance of 10-20\% $X_{\rm s}$ may be reached by either VMS models or lower mass models with additional rotational or convective mixing, also shown in the overlap of solid lines with dashed lines in Fig.\, \ref{fig:XH}.


\section{Conclusions}\label{conclusions}
We present a study of VMS evolution which probes the upper initial mass range for an updated wind recipe accounting for the increased $L/M$ dependency of VMS above the transition point \citep{VG12}. We provide a grid of stellar models at \ZLMC\ for masses 80-1000\Mdot\ with a range of internal mixing prescriptions. Our standard grid of models includes the mass-loss prescription outlined in \cite{Sabh22}, adopted from \cite{vink11}. 

We compare the dynamical masses of both components of the most massive detached binary in the TMBM sample, R144, as a method of analysing the tracers of VMS evolution with comparisons to stellar parameters such as effective temperature, radius and MS age. We provide estimates on the evolutionary history of VMS above 100\Mdot\ and predict the initial masses of stars with current masses in the range 50-100\Mdot. We find important insights from $X_{\rm s}$, particularly for VMS above 300\Mdot\ \citep{best14} and WNh stars shown in Table \ref{tab:VFTS}. 

We find that the differences between increased factors of the base rates of the standard \cite{Vink01} wind recipe and the updated high rates of \cite{Sabh22} accounting for enhanced $L/M$ \corr{exponent}s lead to significant changes in the evolutionary behaviour of VMS, highlighting the importance of the qualitative treatment and implementation of stellar wind recipes. We also provide hydrodynamically-consistent radiative transfer models of the primary component of R144 \corr{(see App. \ref{App1})} to consolidate our interpretation of their empirical properties and confirm their stellar parameters.

We find a mass turnover at $\sim$ 1.6 \,\corr{Myr}, where all models in our standard grid (adopting the $L/M$-dependent wind) overlap and the initial mass is indistinguishable, irrespective of initial mass of the model. This means that of stars with current masses in the range of 50-100\Mdot\ and and age of more than 1.6 \,\corr{Myr}, the initial mass range is unknown. This may have consequences for the upper initial mass\corr{es} of stars in the 30Dor cluster which has been confirmed to have an age of 1-2.5 \,\corr{Myr}, hosting stars up to 200-300\Mdot.

We find a method which can break the degeneracy of the upper initial mass, by utilising the $X_{\rm s}$\, abundance as a proxy for core H-burning evolution due to the chemically homogeneous nature of VMS. In fact, the H clock is an important tool which can decipher which initial mass range the current observations originated from.

Interestingly, we find an important result due to the nature of $X_{\rm s}$ abundance for both chemically-mixed VMS and stars in the range 100-150\Mdot\ with additional mixing via rotation or overshooting. We find that when estimating the initial mass of massive stars there are two main options, firstly the moderate (80-120\Mdot) mass range where stars are not fully-mixed and the $X_{\rm{s}}$ is not a distinguishable factor for values above 30\%. Secondly, the extreme initial mass range of masses above 200\Mdot\ where stars are chemically-homogeneous and the $X_{\rm s}$ fraction can be used as a direct clock for the core evolution. The unique opportunity to break this degeneracy in the moderate and extreme options was shown for R144 where the moderate solution is preferred due to the components $X_{\rm{s}}$\, $=$\, 35-40\%. 

Due to these higher $X_{\rm s}$ abundances of the components of R144, we may more accurately estimate the initial mass range. Alongside the dynamical masses of the components, we can break the degeneracy between extreme and moderate initial mass solutions leading to estimates of 110\Mdot\ and 100\Mdot\ for the primary and secondary respectively, in this case. However, if we observed similar objects with accurate dynamical masses and surface abundances of 0-20\%, we may find an initial mass much higher. 

We also present models with a range of internal mixing efficiencies, with our standard models implementing \aov\,$=$\,0.3 and zero rotation, alongside models which include 40\% critical rotation and models which include \aov\,$=$\,0.5. We find that the $X_{\rm s}$ abundance is a good proxy for the central H abundance in models with $M_{\rm{init}}$\, $>$\,300\Mdot\ due to their CHE nature and strong winds. Models in the range 80-120\Mdot, however, do not show a linear agreement between $X_{\rm c}$ abundance and $X_{\rm s}$\, abundance, particularly when including differing internal mixing efficiencies. Therefore, with dynamical masses and surface abundances well established for R144, we estimate the $M_\mathrm{init}$ to be 110\Mdot\ and 100\Mdot\ for the primary and secondary respectively, which corresponds to the moderate initial mass range \citep[see also][]{Shenar21}.

With uncertainties in $\mathrm{log\corr{_{10}}} \, (T_{\mathrm{eff}}$/K)\, from spectroscopic analysis, it can be speculative to reproduce HRD positions and subsequently predict viable initial mass ranges as we show in Fig.\, \ref{fig:R144_HRD}. Moreover, this can also lead to uncertainties in $\mathrm{log\corr{_{10}}}\;(R/\rm{R}_\odot)$, meaning that neither of these stellar parameters are good constraints if there maintains uncertainties from spectral analysis. On the other hand, dynamical masses from eclipsing binaries --while rare-- provide a key constraint on the evolution of massive stars. However, due to the nature of VMS we can estimate the $X_{\rm c}$ abundance as a function of $X_{\rm s}$\, abundance due to the chemical homogeneity of VMS evolution. 

We find that it is likely that the WNh stars in 30 Dor may have originated from either moderate or extreme initial mass solutions suggesting that the entire initial mass range 100-1000\Mdot\ is possible while having such high luminosities and low $X_{\rm s}$\, abundances. While it remains unknown what exactly is the initial mass of these WNh stars, it is important to consider the wide range of feasible initial masses based on given observations. Therefore, for objects such as VFTS 695 and VFTS 427 we cannot determine if their initial masses were extreme, so initial masses of order 1000\Mdot\ are not excluded.

\section*{Acknowledgements}
JSV and ERH are supported by STFC funding under grant number ST/V000233/1 in the context of the BRIDGCE UK Network. AACS is supported by the Deutsche Forschungsgemeinschaft (DFG - German Research Foundation) in the form of an Emmy Noether Research Group (grant number SA4064/1-1, PI Sander).
\section*{Data Availability}

The data underlying this article will be shared on reasonable request
to the corresponding author.

\bibliographystyle{mnras}
\bibliography{newdiff.bib}

\begin{appendix}
\section{}
\begin{figure*}
    \includegraphics[width = \textwidth]{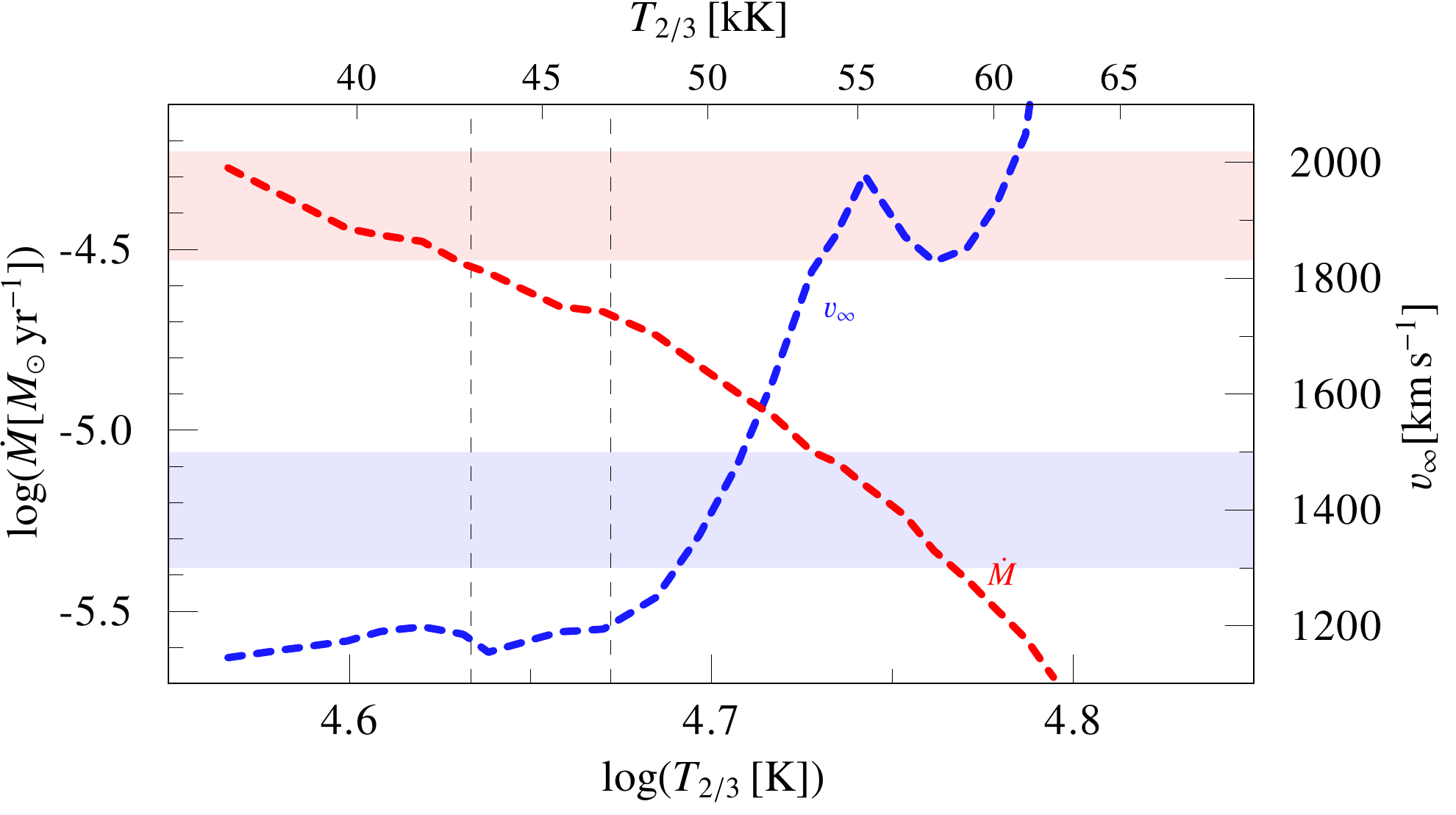}
    \caption{Mass-loss rates and terminal velocities calculated for a range of hydrodynamically (HD) consistent atmosphere models for the primary component of R144 with be $M_\mathrm{dyn}$ $=$ 74\Mdot, calculated with the PoWR code. The red dashed line denotes the mass-loss rate as a function of the effective temperature at \corr{an optical depth of }$\tau$ $=$ 2/3. The blue dashed line denotes the corresponding terminal wind velocity derived by the models. Shaded regions highlight the empirically derived parameters for the R144 primary from \citet{Shenar21} including error margins. The vertical dashed lines mark the empirically derived temperature range for $T_{\mathrm{2/3}}$,  \corr{the temperature calculated at an optical depth of $\tau$ $=$2/3}. }
    \label{fig:hydroAS}
\end{figure*}

\subsection{Hydrodynamically-consistent atmosphere models}\label{App1}
In the empirical analysis of massive stars, it is quite common to obtain mass-loss rates that differ considerably from theoretical recipes. With its orbital \corr{constraints providing accurate dynamical masses}, the massive binary R144 provides a rare opportunity to gauge mass loss in the regime of VMS. For the primary of R144, we calculated a series of hydrodynamically consistent atmosphere models using the technique from \cite{sander17,Sander+2020} around the empirically derived effective temperatures (\corr{calculated at an optical depth of }at $\tau = 2/3$). To keep the number of calculations manageable, we fixed the stellar mass and luminosity to the derived values \corr{(74\Mdot, log$_{\rm{10}}$($L/\Ldot$) $=$ 6.4)} for the primary and assumed a depth-depending clumping with a maximum clumping factor of D $=$ 10. The results presented in Fig.\,\ref{fig:hydroAS} show the trends for the derived mass-loss rates and terminal velocities as a function of the effective temperature of the models, defined at \corr{an optical depth of }$\tau = 2/3$. We obtain a strong temperature dependency for the derived mass-loss rate that drops by an order of magnitude when increasing the temperature from 40kK to 60kK. The derived terminal wind velocities increase, but much more moderate and not strictly monotonic, although this might also be subject to some numerical uncertainty in the models. When compared to the empirically derived wind parameters of \corr{the mass-loss rate }$\dot{M}$ and \corr{terminal velocity }$\varv_\infty$, \corr{calculated from stellar atmosphere models of the observed spectra}, indicated by shaded areas in Fig.\,\ref{fig:hydroAS}, it is evident that the full combination of derived stellar and wind parameters for the R144 primary is in line with the theoretical modelling results, \corr{ within $\sim$0.2dex}. For the empirical temperature, the calculated $\dot{M}$ and $\varv_\infty$ are both slightly too low \corr{(by $\sim$0.2dex)} and any temperature shift that would lower one discrepancy would increase the other.

Given that we did not account for any uncertainty in the mass and luminosity in our calculation, it is likely that already a slightly higher \corr{value of} $L/M$ would be able to fix this small discrepancy. However, a multi-parameter study of such models is not only beyond the scope of this paper, but also numerically very costly and thus has to be postponed to a separate, later study. 

Nonetheless, our atmosphere modelling efforts underline that the empirically derived mass-loss rates are -- within a reasonable error margin -- backed by theoretical wind driving studies. This implies that the wind mass agrees with the dynamical mass giving confidence to present day mass estimates.

\end{appendix}

\end{document}